\documentclass[manuscript]{aastex}

\newcommand{\Hipp}{{\it Hipparcos}}        

\newcommand{\HST}{{\it HST}}

\newcommand{\Teff}{T_{\rm eff}}
\newcommand{\Mjup}{M_{\rm Jup}}

\newcommand{\kms}{{\>\rm km\>s^{-1}}}
\newcommand{\msun}{M_\sun}

\shorttitle{HST Astrometry of Procyon}
\shortauthors{Bond et al.}

\begin{document}

\title{{\em Hubble Space Telescope\/} Astrometry of the Procyon
System\altaffilmark{1} }

\author{
Howard E. Bond,\altaffilmark{2,3}
Ronald L. Gilliland,\altaffilmark{3,4}
Gail H. Schaefer,\altaffilmark{5}
Pierre Demarque,\altaffilmark{6}
Terrence~M. Girard,\altaffilmark{6}
Jay B. Holberg,\altaffilmark{7}
Donald Gudehus,\altaffilmark{8}
Brian D. Mason,\altaffilmark{9}
Vera~Kozhurina-Platais,\altaffilmark{3}
Matthew R. Burleigh,\altaffilmark{10}
Martin A. Barstow,\altaffilmark{10}
and
Edmund~P.~Nelan\altaffilmark{3}
}

\altaffiltext{1} {Based on observations with the NASA/ESA {\it Hubble Space
Telescope\/} obtained at the Space Telescope Science Institute, and from the
Mikulski Archive for Space Telescopes at STScI, which are operated by the
Association of Universities for Research in Astronomy, Inc., under NASA contract
NAS5-26555 }

\altaffiltext{2}
{Department of Astronomy \& Astrophysics, Pennsylvania State
University, University Park, PA 16802, USA; heb11@psu.edu}

\altaffiltext{3}
{Space Telescope Science Institute, 
3700 San Martin Dr.,
Baltimore, MD 21218, USA}

\altaffiltext{4}
{Center for Exoplanets and Habitable Worlds,
Department of Astronomy \& Astrophysics, Pennsylvania State
University, University Park, PA 16802, USA}

\altaffiltext{5}
{The CHARA Array of Georgia State University, Mount Wilson Observatory,
Mount Wilson, CA 91023, USA}

\altaffiltext{6}
{Department of Astronomy, Yale University, Box 208101, New Haven, CT 06520, USA}

\altaffiltext{7} 
{Lunar \& Planetary Laboratory, University of Arizona, 1541 E. University
Blvd., Tucson, AZ 85721, USA}

\altaffiltext{8}
{Department of Physics \& Astronomy, Georgia State University, Atlanta, GA
30303, USA}

\altaffiltext{9}
{U.S. Naval Observatory, 3450 Massachusetts Ave., Washington, DC 20392, USA}

\altaffiltext{10}
{Department of Physics \& Astronomy, University of Leicester, Leicester LE1
7RH, UK}

\begin{abstract}

The nearby star Procyon is a visual binary containing the F5~IV-V subgiant
Procyon~A, orbited in a 40.84~yr period by the faint DQZ white dwarf
Procyon~B\null. Using images obtained over two decades with the {\it Hubble
Space Telescope}, and historical measurements back to the 19th century, we have
determined precise orbital elements. Combined with measurements of the parallax
and the motion of the A component, these elements yield dynamical masses of
$1.478\pm0.012\,M_\odot$ and $0.592\pm0.006\,M_\odot$ for A and B, respectively.

The mass of Procyon A agrees well with theoretical predictions based on
asteroseismology and its temperature and luminosity.  Use of a standard
core-overshoot model agrees best for a surprisingly high amount of core
overshoot. Under these modeling assumptions, Procyon~A's age is
$\sim$2.7~Gyr.  

Procyon~B's location in the H-R diagram is in excellent agreement with
theoretical cooling tracks for white dwarfs of its dynamical mass. Its position
in the mass-radius plane is also consistent with theory, assuming a
carbon-oxygen core and a helium-dominated atmosphere. Its progenitor's mass was
1.9--$2.2\,M_\odot$, depending on its amount of core overshoot. 

Several astrophysical puzzles remain. In the progenitor system, the stars at
periastron were separated by only $\sim$5~AU, which might have led to tidal
interactions and even mass transfer; yet there is no direct evidence that these
have occurred. Moreover the orbital eccentricity has remained high ($\sim$0.40).
The mass of Procyon~B is somewhat lower than anticipated from the
initial-to-final-mass relation seen in open clusters. The presence of heavy
elements in its atmosphere requires ongoing accretion, but the place of origin
is uncertain.

\end{abstract}

\keywords{astrometry --- stars: binaries: visual --- stars: fundamental
parameters --- stars: individual (Procyon) --- stars: white dwarfs}

\section{Introduction}

An early triumph of positional astronomy was the discovery of astrometric
perturbations of the motions of Procyon and Sirius by Bessel (1844), who
attributed them to the gravitational influence of unseen companions. In 1862,
Sirius~B was seen by Alvan G. Clark,\footnote{Clark's discovery was reported,
and confirmed, by Bond (1862).} verifying Bessel's supposition. Procyon
($\alpha$~Canis Minoris) proved more recalcitrant: many visual observers
attempted during the rest of the 19th century to detect its companion, but
without success---even though by this time it was known from the astrometry that
the orbital period is about 40~yr (Auwers 1873), and even approximately where
the companion should be located. Struve (1874) reported an extremely faint
companion of Procyon, but his claim was subsequently withdrawn and conceded to
be spurious. Extensive attempts at the U.S. Naval Observatory (USNO) in 1874 and
1876 also failed (Davis 1876). Finally, more than five decades after Bessel's
announcement, the faint companion of Procyon was seen visually by Schaeberle
(1896) at the Lick Observatory 36-inch refractor. The history of the
19th-century searches for Procyon~B was colorfully summarized by See
(1898)---who called the Procyon system ``the most magnificent which astronomical
observation has yet disclosed''---and later by Spencer Jones (1928).

The companion, Procyon~B (WD 0736+053), with a visual magnitude of 10.82
(Provencal et al.\ 2002, hereafter P02), is 10.5~mag fainter than its primary, and is never
separated by more than $\sim$$5''$ from Procyon~A\null. Visual detection of
Procyon~B is thus notoriously difficult. For example, Schaeberle (1897) remarked
that it is ``useless to look for either of these companions [of Procyon and
Sirius] with the 36-inch telescope when the seeing is not good.'' At the Yerkes
Observatory, the eminent visual observer Barnard (1913) noted that ``it requires
the best conditions to even see it.'' Charles Worley, double-star observer at
the USNO, asserted to two of us, more than two decades ago, that he was the only
living astronomer who had seen Procyon~B with his own eye.

Nevertheless, visual measurements of the separation and position angle of
Procyon~B slowly accumulated over the first third of the 20th century, along
with absolute astrometry and radial-velocity (RV) measurements of
Procyon~A\null. Comprehensive analyses of these data were made by Spencer Jones
(1928) and Strand (1951, hereafter S51), yielding dynamical masses for the two
components ($M_A$ and $M_B$) of 1.24 and $0.39\,\msun$, and 1.74 and
$0.63\,\msun$, respectively. Procyon~A is a slightly evolved subgiant of
spectral type F5~IV-V (e.g., Gray et al.\ 2001). The very low luminosity of B
relative to its mass indicated that it must be a white dwarf (WD), as deduced by
Kuiper (1937). In fact, it is the third brightest WD in the sky after Sirius~B
and $\rm o^2$~Eridani~B---but it was not possible to obtain its spectrum until
the advent of the {\it Hubble Space Telescope\/} (\HST\/)\null. Using \HST\/
spectra, P02 showed that Procyon~B has a spectral type of DQZ (i.e., a WD whose
spectrum is devoid of hydrogen and shows features due to carbon, magnesium, and
iron). 

Irwin et al.\ (1992, hereafter I92) discussed precise RVs of Procyon~A and
re-analyzed the published astrometry. Their fit to these combined data yielded
masses of 1.75 and $0.62\,\msun$, very close to the values reported by S51
four decades earlier. However, it had been pointed out by several authors (e.g.,
Hartmann, Garrison, \& Katz 1975; Steffen 1985; Demarque \& Guenther 1988; 
Guenther \& Demarque 1993) that stellar models that match the observed
luminosity and effective temperature of Procyon~A require its mass to be close
to $1.50\,\msun$. I92 noted that the discrepancy could be resolved if the
semimajor axis of the relative orbit of B around A (at that time still based
entirely on visual observations) were systematically too large by
$\sim$$0\farcs2$.

Girard et al.\ (2000, hereafter G00) measured some 600 photographic exposures,
spanning 83 years, to redetermine the parallax and astrometric motion of
Procyon~A\null. To this analysis they added measurements of the A--B separation
made in 1995 from a ground-based near-infrared coronagraphic observation, and
from an observation obtained with the Wide Field Planetary Camera~2 (WFPC2) on
\HST\null. This study resulted in a substantially reduced dynamical mass of
$1.497\pm0.036\,\msun$ for Procyon~A, along with $0.602\pm0.015\,\msun$ for the
WD\null. The discrepancy with stellar theory thus appeared to have been removed.
(It proved to be correct, as anticipated by I92, that the visual determination
of the relative semi-major axis had been too large by about
$0\farcs2$---although this conclusion was still based almost entirely on the
single \HST\/ observation.)

However, the mass of Procyon~A remained a subject of debate over the years
following the G00 publication, with several investigators, both theoretical and
observational, advocating even lower values. Allende Prieto et al.\ (2002),
adopting a smaller parallax and giving higher weight to the 1995 WFPC2
separation measurement, found a mass of $1.42\pm0.06\,\msun$\null. Kervella et
al.\ (2004), on the basis of asteroseismic data and an interferometric
measurement of the angular diameter of Procyon~A, argued that its mass is as low
as $\sim$$1.4\,\msun$\null. Relatively low masses for both components were also
deduced by Gatewood \& Han (2006), who found $1.431\pm0.034$ and
$0.578\pm0.014\,\msun$ in an analysis that included their astrometric data on
Procyon~A from the Allegheny Observatory Multichannel Astrometric Photometer
(MAP)\null. But Guenther et al.\ (2008) emphasized again that masses this low
are difficult to reconcile with stellar models, which require a mass close to
$1.5\,\msun$\null. Recent discussions of these issues, and further references,
are given by Chiavassa et al.\ (2012) and Liebert et al.\ (2013, hereafter
L13).  

%

As we have recounted, visual measurements of the orbit of Procyon~B are subject
to large systematic errors; and the pair is likewise difficult even for modern
ground-based instrumentation. But in sharp contrast to ground-based
observations, the Procyon system is easily resolved and measured in
appropriately exposed \HST\/ images. Because of the importance of Procyon~A as a
fundamental calibrator of stellar physics on or near the main sequence, and of
the mass of Procyon~B for our understanding of WDs, our team began a program of
regular \HST\/ imaging of the system. Our aims were to obtain dynamical masses
of both stars with the highest possible precision, and an accuracy limited only
by factors such as the absolute parallax of the system. Moreover, precise
relative astrometry of the binary can place limits on the presence of third
bodies in the system, down to planetary masses. Our project began with WFPC2 in
1997 November, and we observed the Procyon system at a total of 11 epochs
through 2007 October. We then continued the program with the Wide Field Camera~3
(WFC3), following its installation in place of WFPC2 during the 2009 Servicing
Mission. Our WFC3 images were obtained at five epochs between 2010 February and
2014 September. In addition, the \HST\/ archive contains the 1995 observations
mentioned above, and two more observations in 1997, for a grand total of 19
epochs between 1995 and 2014, covering almost half of the orbital period.


In this paper, we present the precise relative astrometry of the binary that
results from the \HST\/ observations, determine the elements of the visual
orbit, and derive updated dynamical masses for both stars. We then discuss the
astrophysical implications for the two components, and place limits on the
presence of third bodies in the system.

%
%
%
%


\section{{\em HST\/} Observations}

Procyon~A, at $V=0.34$ (Johnson \& Morgan 1953), is the eighth brightest star in
the sky. The companion WD is fainter at visual wavelengths by nearly a factor of
16,000. Astrometry of this binary, even with \HST, therefore presents an
observational challenge. It is too bright for astrometric observations with the
Fine Guidance Sensors (FGS)\null. With WFPC2, whose shortest allowable exposure
time was 0.11~s, it was possible to obtain unsaturated images of Procyon~A only
by using a filter bandpass, F218W, located at the short-wavelength extreme of
the CCD detector sensitivity.  The approach we adopted was therefore to take a
short, unsaturated exposure on Procyon~A, and then, without moving the
telescope, a second exposure long enough to detect Procyon~B with good
signal-to-noise ratio (SNR)\null. This procedure was then repeated at several
additional dithered positions (typically for a total of five exposure pairs)
during the \HST\/ orbit. All of our WFPC2 observations were taken with the F218W
ultraviolet filter, as were the 1995 archival frames mentioned above. A further
advantage of observing in the ultraviolet is that the contrast between the stars
is reduced to a factor of about 2,600, due to the WD being somewhat hotter than
Procyon~A\null. In addition to these frames, the archive contains images taken
at two epochs in 1997, using the F1042M filter at the long-wavelength end of the
WFPC2 sensitivity.\footnote{There are also limited archival observations of
Procyon obtained with WF/PC-1, NICMOS, and STIS, and with other WFPC2 filters
than the ones we used, but we judged these unlikely to contribute additional
useful astrometric data.}  Procyon~A is saturated in these frames. For all of
the WFPC2 observations, Procyon was placed near the center of the Planetary
Camera (PC) CCD, which has a plate scale of $0\farcs0454 \, \rm pixel^{-1}$. We
requested telescope roll angles such that Procyon~B would not lie near the
diffraction spikes or charge bleeding of the bright component. 


When the much more sensitive WFC3 was installed in place of WFPC2, it became
impossible to obtain unsaturated images of Procyon~A in any of the available
filters, even using the shortest allowable exposure time of 0.5~s. Our approach
was instead to take fairly deep dithered images, yielding a good SNR on
Procyon~B, and to locate the centroid of Procyon~A using features (primarily the
diffraction spikes) outside the saturated center of its image. For the WFC3
images, we chose the UVIS channel (plate scale $0\farcs0396 \, \rm pixel^{-1}$)
and its near-infrared narrow-band F953N filter. Apart from the low system
throughput in this filter---desirable for this particular application---an
advantage of the long wavelength was the resolution of the diffraction spikes
into a triple structure (due to the first Airy ring), whereas these features are
blended into a single blurred spike at shorter wavelengths. (On the other hand,
the contrast between the stars in this filter is nearly a factor of 20,000.) For
the WFC3 imaging, Procyon was placed near the center of a $512\times512$
subarray (in order to reduce data volume and improve observing efficiency).

Observing logs for the WFPC2 and WFC3 data are presented in Tables~1 and~2.


\section{{\em HST} Astrometric Analyses}

For the measurements of separation and position angle for the Procyon
system, we have three distinct sets of \HST\/ data, each requiring different
astrometric analysis techniques. These are (1)~the (mostly) unsaturated images
obtained with WFPC2 and the F218W filter; (2)~a set of WFPC2 frames in the
F1042M filter, in which Procyon~A is saturated; and (3)~the WFC3 images in
F953N, in which Procyon~A is also saturated.

\subsection{WFPC2 Images in F218W}


Figure~1 illustrates a typical pairing of a short unsaturated exposure in F218W
for astrometry of Procyon~A, and a much longer  exposure at the same pointing
used to analyze Procyon~B\null. The inset in the center shows Procyon~A from a
0.11-s exposure, superposed on a 100-s frame in which the WD is easily detected
in spite of the neighboring, grossly overexposed image of the primary star.


In our first visit in 1997, we were too aggressive  in choosing an exposure time
for A of 0.14~s, resulting in its image containing saturated pixels in most of
the exposures. Fortunately, due to dithering, there were two short-exposure
frames in which A remained unsaturated.  Even at the reduced 0.12~s used in the
1998 visit, two of the short exposures were again saturated for Procyon~A\null. 
For the remainder of the WFPC2 observations, we set the short exposures to the
WFPC2 minimum of 0.11~s, and none of them were saturated.

In all cases we used the individual {\tt c0m.fits} images from the archive
pipeline for the astrometric analysis. These frames have bias subtraction and
flat-fielding applied, but do not include any cosmic-ray removal, geometric
correction, or drizzle processing. Each short- and long-exposure pair was taken
at a different dither position, using fractional-pixel offsets to sample the
point-spread function (PSF), plus shifts of a few integer pixels to average out
the impact of detector defects (such as hot pixels). We checked for discrepant
measurements due to cosmic-ray impacts, but found no cases where they had caused
a problem in our relatively short integrations.

In the analysis of the 1995 WFPC2 observation by G00, the relative
positions of Procyon~A and
B were determined by cross-correlation of the short-exposure image of A with the
long-exposure image of B\null. The uncertainty for this measurement was at a
level of about $\pm$0.2 pixels ($\sim$$0\farcs009$). However, the accumulation
of WFPC2 data from several programs, including ours, which used F218W between
1994 and 2009, makes it possible to apply a more precise astrometric analysis
based on PSF fitting. (We did try the original cross-correlation approach for
our WFPC2 data, but found that the errors were about 50\% greater than those
based on PSF fitting.)

PSF fitting is based on an empirically derived, over-sampled representation of
the image structure, obtained by combining numerous high-SNR exposures taken at
many independent pointings. As indicated in Table~1, we have 53 unsaturated
exposures of 0.11 to 0.14~s on Procyon~A\null. In addition, we included 96
archival observations of the standard star Grw~+70$^\circ$5824, obtained for
WFPC2 F218W photometric calibrations between 1994 and 2009, and two frames of
the standard star BD~+17$^\circ$4708 from 2004. That made a total of 151
well-exposed F218W images of stars having colors similar to Procyon~A and B,
positioned near the center of the PC chip, for input to the PSF determination.

%

The approach we used is described in eqs.~(2) and (3) of Gilliland et al.\
(1999).  A uniform spatial grid on a scale finer than the native pixel size is
first defined. (In this case, we chose a factor of 50 finer than the input
scale, because it allowed re-use of existing codes developed for {\em Kepler\/}
analyses, but the results are insensitive to the exact choice.)  

The individual pixel values, after normalization of all inputs to unit volume,
are accumulated into a weighted sum at each over-sampled grid point, using a
Gaussian weighting based on separations of each input pixel from the
accumulation grid.  The width of the  Gaussian weighting function is a free
parameter.  Adoption of too small a Gaussian width results in over-fitting the
data, while too wide a weighting function suppresses available
resolution in the resulting PSF\null.  We used a Gaussian weighting width  of
0.416~pixels full width half maximum (FWHM), which minimized the scatter in
measurements at the same epoch.

Developing the over-sampled PSF requires precise knowledge of the relative
centering of each input image. The solution is therefore iterative, since
precise relative positions are best determined through fitting the over-sampled
PSF to individual images.  Fortunately a simple first-moment estimate of image
positions for all inputs is accurate enough to start a rapidly convergent
iterative cycle of determining an over-sampled PSF, revising the image
positions, and recalculating the PSF.

With the PSF defined, we then obtained the relative positions of individual
images of Procyon~A and B by fitting a bi-cubic interpolation function (Press et
al.\ 1992) to the PSF, and then employed a non-linear least-squares fit
(Bevington 1969) to determine the relative $x, y$ centers of both stars. These
fits used the central 21 pixels ($5 \times 5$ box without corners), after
experiments showed that smaller or larger fit domains performed marginally less
well.

Procyon~B lies in the extended wings of the PSF of Procyon~A\null. Simulations
indicated that these wings shift the measured position of B by less than
$0\farcs001$ at times of greatest separation early in the WFPC2 series,
increasing to $0\farcs003$--$0\farcs004$ near closest approach. We therefore
derived a deep PSF by stacking all of the strongly saturated Procyon A images,
which we then subtracted before performing PSF fits for the position of B\null. 

WFPC2 had significant geometric distortion, due both to the camera optics and a
manufacturing defect in the CCDs. We applied geometric-correction terms for the
optical distortions, and the ``34th-row'' detector defect, from Gonzaga \&
Biretta (2010). The geometrically corrected $x,y$ positions were then converted
to angular units using a F218W plate scale of $0{\farcs}045437\rm\,pixel^{-1}$,
with a nominal fractional error of $\pm$0.0003, adjusted slightly for
differential velocity aberration (using the {\tt VAFACTOR} keyword in the image
headers), all as described by Gonzaga \& Biretta.\footnote{The WFPC2 CCDs also
suffered from charge-transfer inefficiency (CTI), which increased with time over
the years spent in the space environment. At low light levels, CTI can cause
systematic shifts in the centroid locations of stellar images. However, in the
case of our Procyon observations, involving bright stars and strong background
light, the CTI effects on the astrometry are very small (e.g., Goudfrooij et
al.\ 2006), and we did not make any corrections for them.}



Because of our technique of short exposures followed by long exposures, our
measurements of the separation and position angle are subject to a systematic
offset due to telescope pointing drift (which occurs even when the telescope is
locked on guide stars). Gilliland (2005) showed that drifts of
$0\farcs010$--$0\farcs015$ are typical during \HST\/ orbits, which translates to
about $0{\farcs}0025$ for a pair of A, B exposures taken over about 1/5th of an
\HST\/ visibility period.


Since our images only show the two components of Procyon, we are unable to
establish firmly that the pointing drift was always present, and if so what its
direction was; but the effect of such drifts is likely to dominate over other
terms (such as influence of differing stellar colors or changing telescope focus
on the PSF, plate-scale changes due to telescope ``breathing,''  residual
contamination from component A, etc.).   We have therefore estimated the errors
of the average positions at each epoch by combining in quadrature the standard
error based on the observed measurement scatter with a systematic term of
$\pm$$0{\farcs}0025$ for telescope drift. (Although drift is systematic within a
single \HST\/ visit, a range of different telescope orientations was used across
the different epochs, so it is appropriate to treat drift error as a random
term.)

%


Lastly, we determined the absolute J2000 position angle of B relative to A,
using the {\tt ORIENTAT} keyword in the image headers, which gives the
orientation on the sky of the image $y$ axis. The error on position angle
includes two terms. The first  arises from the errors of derived $x, y$
positions of A and B, estimated as described above. This term in position angle
will be inversely proportional to the lever arm provided by the changing
separation of A and B during their orbit. A second term arises from
uncertainties in the absolute {\em HST\/} roll angle. We assume a $1''$ error on
guide-star positions, observed with the FGS over a $\sim$$1000''$ baseline,
which translates to an angular error of $\pm$$0\fdg028$. We combine these two
uncertainties in quadrature. The final astrometric results from the WFPC2 F218W
images are given in Table~3, lines 1 and 4 through~14.


\subsection{WFPC2 Images in F1042M}


The \HST\/ archive contains images of Procyon in the WFPC2 F1042M filter 
obtained at two epochs in 1997. The primary aim of these observations was a
search for faint companions of 23 nearby stars, including Procyon, but no new
companions were found by the proposing team (Schroeder et al.\ 2000). For our
astrometric measurements, these frames raised the challenge that Procyon~B is
nicely exposed, but the core of the image of A is strongly saturated. This
forced us to develop a means of using the outer portions of the PSF of
Procyon~A, where the spatial information is dominated by the diffraction spikes,
to obtain its centroid location.

We faced a similar problem in our complementary \HST\/ program on the Sirius
system, to be discussed in a separate forthcoming paper. Our WFPC2 images of
Sirius were taken in the same F1042M filter, and are likewise saturated for
Sirius~A\null. In order to test methods for centroiding saturated images, we
carried out a calibration program (Program ID: CAL/WFPC2-11509) on the star
109~Virginis (spectral type A0~V, $V=3.73$). This star has a color similar to
that of Sirius, and not extremely different from Procyon. It is sufficiently
faint that unsaturated images in F1042M can be obtained in short exposures
(0.23~s), along with saturated images from longer integrations (600~s). We
obtained a set of three dithered pairs of short and long exposures on this
star.\footnote{Janson et al.\ (2011) reported ground-based detection in the
near-IR ($1.58\,\mu$m) of a companion to 109~Vir, at a separation of $0\farcs57$
and 6.04~mag fainter, epoch 2010.52. This object was also detected on 2009.27
and 2010.32 by L. C. Roberts et al.\ (private communication), at a similar
position, and 5.4~mag fainter at $2.1\,\mu$m. They did not detect it at
$1.25\,\mu$m. This location lies within the saturated pixels of our long \HST\/
exposures. In our short exposures at $1.04\,\mu$m no companion is seen at the
positions given by Janson et al.\ and Roberts et al.\ (nor anywhere else in the
field), even though a 6-mag-fainter star would be readily detectable. This
suggests that the companion is extremely red.}



We initially considered an approach for astrometric analysis of the saturated
images of Procyon in which we would develop an over-sampled PSF for the
unsaturated regions of the images, including especially the diffraction spikes.
As in the case of the unsaturated F218W images described above, development of
the PSF for saturated F1042M images requires an ensemble of input data. Table~4
lists the images of Sirius and 109~Vir that we used (in addition to the two sets
of Procyon~A F1042M images listed in Table~1). 

However, we found that the appearance of the diffraction spikes is unstable.
Figure~2 shows examples of the variable spike structures in saturated images of
Sirius and Procyon in the F1042M filter. The intensities of the spikes vary by
large amounts as functions of distance from the center of the stellar image in a
quasi-periodic fashion, which does not reproduce well from epoch to epoch. The
structure of these intensity variations appears to depend strongly on small
differences in the location of the star in the field of view. Therefore we did
not see any straightforward means of defining an over-sampled PSF for the
unsaturated outer regions of the deep exposures.

We instead adopted an alternative approach of fitting straight lines to the
diffraction spikes, and determining the image centroid from their intersection
point.  Our procedure was to estimate the location of the pixel nearest the
center of the saturated image, and then search inwards toward this point along
each of the four diffraction spikes until the first saturated pixel was
encountered. From that pixel outward, we calculated the sums of intensities
along each diffraction-spike axis and along the two neighboring parallel axes one
pixel away on either side. These sums were accumulated in two sequential
segments, each 30 pixels long, for a total length of 60~pixels. Then a parabola
was fit to the three sums; the peak of this parabola marked the location of the
diffraction-spike axis in the direction orthogonal to the spike. We found good
consistency between results from the first and second 30-pixel segments along
the spikes, and thus combined them to form a single center for each of the four
spikes. The intersection point of the lines connecting the symmetric
diffraction-spike centers then defined the stellar centroid.

Application of this approach to the three pairs of calibration frames on 109~Vir
gave mean offsets between the spike intersection point and the PSF-derived
centroid of $-0.004 \pm 0.002$ pixels in $x$, and $+0.044 \pm 0.040$ pixels in
$y$ (or, in arcseconds, $-0\farcs0002\pm0\farcs0001$ and
$+0\farcs0020\pm0\farcs0018$, respectively; the errors are estimated from the
scatter among the three measurements). We simply adopted these as (small)
corrections to be added to the A-component positions from the long exposures.


To determine the centroids of Procyon~B from the images, we followed the
approach used for the F218W data. That is, an over-sampled PSF was first
derived, using the same Gaussian weighting approach.  We used 69 individual
inputs, listed in Tables~1 and 4, consisting of all deep exposures of Procyon~B
and Sirius~B in F1042M, plus the three unsaturated exposures on 109~Vir.
Background removal was done by taking a median of an annulus from 13 to 23
pixels out from B, then subtracting it to remove the pedestal due to the wings
of A\null.  

Due to a MgF$_2$ lens immediately in front of the CCDs, there is a weak
dependence of the WFPC2 plate scale on wavelength. A plate scale for F1042M
images is not provided in Gonzaga \& Biretta (2010). However, by plotting the
plate scales listed by Gonzaga \& Biretta against the index of refraction of
MgF$_2$ at the effective wavelength of each filter, we found a tight, linear
correlation. Only a slight extrapolation to the wavelength of F1042M was needed
to estimate its plate scale. We adopted a relative plate scale of 1.00048
compared to the fiducial F555W value, for a net of $0\farcs045577\,\rm
pixel^{-1}$ for the PC chip. 

Typical WFPC2 F1042M visits consisted of five or six exposures of about 8~s, and
another five or six of about 60~s. Thus there are two distinct clumps of medium
and long integrations. To see if there was a dependence of the astrometric
results on exposure time, we compiled means and scatters within the medium and
long blocks separately for each epoch. The average difference in measured
separations from the medium- and long-exposure sets was an inconsequential
0{\farcs}0007. The average scatter of separations was 0{\farcs}0026 within the
medium exposures, and 0{\farcs}0035 for the long exposures. We concluded that we
could safely combine the results from all of the exposures within each epoch.

The remainder of the F1042M astrometric analysis proceeded as described above
for the F218W images. However, the Procyon astrometry is not directly affected
by {\em HST\/} pointing drift, since A and B are measured on the same frames. On
the other hand, our results hinge on the single-orbit calibration using three
pairs of short and  long 109~Vir exposures, which {\it were\/} subject to the
drift error. The canonical drift allowance of $0\farcs010$--$0\farcs015$ per
\HST\/ orbit visibility translates in this case into a potential systematic
error of $\sim$$0\farcs004$. We applied this value in quadrature with the the
standard error based on random scatter. The astrometric results from the WFPC2
F1042M images are given in Table~3, lines 2 and~3.


\subsection{WFC3 Images in F953N}

The WFC3 observations of Procyon in F953N are similar to those in WFPC2 F1042M:
Procyon~A is saturated in all images. We adopted a similar approach for the
analysis, beginning by assembling a set of images for PSF determination and a
study of the use of diffraction spikes for centroiding. In addition to the WFC3
F953N images of Procyon, listed in Table~2, we have observed Sirius in this WFC3
filter in our complementary program on that binary. And we likewise carried out
calibration observations (Program ID: GO-12598), in which we obtained both
saturated and unsaturated WFC3 frames in F953N of the Pleiades main-sequence
star HD~23886 (spectral type A3~V, $V=8.01$). Table~5 lists the images of Sirius
and HD~23886 that we used for these studies. 

The data for all three targets were acquired using four-point dithering with the
{\tt WFC3-UVIS-DITHER-BOX} pattern.  In most cases repeats were used at each
setting within the pattern, providing 32 total exposures during each \HST\/
visit for Procyon, 28 for Sirius, and 8 for HD~23886.

The HD~23886 calibration observations, and most of our Procyon observations,
used Chip~2 of the WFC3 camera, with {\tt UVIS2-C512C-SUB}, the
$512\times512$-pixel subarray nearest the Amp~C readout. This subarray has been
shown (Gilliland et al.\ 2010) to be the best behaved for photometry near and
beyond saturation.  The Sirius observations all used the larger {\tt
UVIS2-C1K1C-SUB} $1024\times1024$ subarray, also in Chip~2.  However, our first
WFC3 visit for Procyon in 2010 was obtained using the corresponding Chip~1
subarray. Lacking any supporting calibration observations for this chip, we have
omitted the 2010 data from our analysis.

For the WFC3 astrometric analysis, we use the default drizzle-combined {\tt
drz.fits} images from the archive pipeline. These frames are created by
combining the individual dithered exposures, and are fully processed to
bias-subtracted, flat-fielded, and geometrically corrected images with cosmic
rays removed. 

The two left-hand panels in Figure~3 show two representative images of Procyon
from our WFC3 F953N observations. In contrast to the WFPC2 frames in F1042M,
the diffraction spikes have a smooth fall-off in intensity with radius, without
any quasi-periodic fluctuations. Moreover, the image structure appears to be
consistent over all of the epochs. 

We therefore developed a deep PSF, using all of the F953N exposures in Tables~2
and 5, except for the short unsaturated exposure on HD~23886. This PSF extends
out to a large enough radius always to cover the location of Procyon~B; a small
region around B was set to zero weight in each individual exposure contributing
to the deep PSF\null. For the weighting we adopted a Gaussian weighting FWHM of
0.832~pixels.  With the deep PSF determined, we then subtracted it before
using data on B (both for development of the unsaturated, core PSF, and for the
subsequent centroiding of B).

The two right-hand panels in Figure~3 show the result of subtracting the
best-fit deep PSF from two individual images. A region around the primary-star
charge  bleeds is not  handled well, but this is inconsequential. Apart from
this, within the diffraction spikes, and generally for fine structures in the
PSF from A, the  subtraction effectively removes 85--90\% of the flux.

As shown by the red boxes in Figure~3, we used only relatively small regions
containing high-SNR but unsaturated point-like structures, in each of the four
diffraction spikes, for the PSF fitting.  The signal within these boxes exceeds
that from B by over an order of magnitude.  

Positions of B in all cases were determined using the PSF-fitting approach
adopted for F218W\null. To create the PSF for this purpose, we started by
stacking the 25 drizzled images of Procyon~B, Sirius~B, and HD~23886
(unsaturated). For the Procyon~B and Sirius B inputs, the underlying light from
A was first subtracted, using the deep over-sampled PSF of A.

Since we used two independent methods for fitting A and B in the same frames,
and the PSFs do not have absolute centroids, it is important to apply a
calibration using the images of HD~23886. We found that corrections of +0.468
pixels in $x$, and +0.449~pixels in $y$, needed to be added to the A-centroid
technique results to bring them into alignment with the B technique. This leaves
the possibility of telescope drift during the calibration observations 
unaccounted for. Fortunately, however, inspection of the HD~23886 images showed
that it has a (previously unknown) faint companion, offset by $\sim$20 pixels in
$x$, and $\sim$5 in $y$, from the bright star, which is detected in both the
short and long exposures.\footnote{The companion of HD~23886 that we detected at
epoch 2012.1290 is at separation $0\farcs804\pm0\farcs005$ and J2000 P.A.
$228\fdg94\pm0\fdg35$, and is 5.6~mag fainter in F953N.} This allows a direct
correction for \HST\/ pointing drift (particularly valuable in this case because
one of the short\slash long pairs was split across two spacecraft orbits).
Monitoring the faint-star position with unsaturated PSF fits in both the long
and short exposures indicated a correction for drift of +0.026 and $-0.018$
pixels in $x$ and $y$, respectively, in the above sense. Thus the net
calibration zero-point corrections are +0.494 and +0.431~pixels in $x$ and $y$.
Since the companion is faint in the short exposures, the precision is rather
low, and we adopt a systematic error term of $\pm$0{\farcs}004 for the WFC3
astrometry within epochs. This was added in quadrature as a random term since
visits are at effectively random orientations.

Although the pipeline images are geometrically corrected, at the time of our
initial analyses the geometric distortion in F953N had not been calibrated as
well as for the more frequently used WFC3 filters. In particular, it was not
included in the study of WFC3 plate scales by Kozhurina-Platais (2014). 
However, a search of the \HST\/ archive yielded a set of frames in F953N of the
cluster $\omega$~Centauri. We performed a new analysis of these images,
generating new geometric-distortion calibration reference files paralleling
those in the work just cited for other filters. After these files were
incorporated into the calibration database at STScI, we retrieved the data
again, and the results presented here make use of the new calibrations. The
final plate scale adopted in the pipeline reductions for F953N is
$0\farcs03962\,\rm pixel^{-1}$. The results from this method are shown in
Table~3 as the ``WFC3 F953N PSF fit'' entries.

Having developed the alternate technique of fitting straight lines to the
diffraction spikes, and then taking the intersection of these lines as the
centroid of component A for the F1042M data, we also applied this technique to
the WFC3 F953N data. We again used the short and long exposures on HD~23886 to
calibrate the offset between the spike-determined position of A and the
PSF-determined position of B\null. In this case, the corrections are +0.055 and
+0.102 pixels, to be applied to the position of A\null. The results from this
method are labelled in Table~3 as ``WFC3 F953N spike fit'' values. The largest
absolute difference in A-B separation between the two methods over the four
epochs is 5 mas, with a mean of 1.6 mas, and a standard deviation of 2.9 mas.
This suggests that the two techniques yield comparable results, with differences
between them consistent with the stated error bars. We therefore averaged the
results, and show them at the bottom of Table~3, labelled ``WFC3 F953N
average.'' 


\section{Orbital Solution}

\subsection{Compilation of Ground-based Measurements}

Our \HST\/ measurements of the Procyon system are extremely precise, compared to
ground-based data, but they cover less than half of only one orbital period.
Thus the historical ground-based data are important in constraining the orbital
elements, especially the period. The available visual observations of Procyon,
from 1896 to 1932, were assembled by S51 (his Table~8). Since 1932, according to
the Washington Double Star (WDS) Catalog maintained at the USNO, there have been
only nine further published measurements of Procyon. Three of these measurements
are from \HST\/ observations in 1995 and 1997 (G00; Schroeder et al.\ 2000), now
superseded by our present results, and leaving only six new published
ground-based observations since 1932.



The early observers would often report measurements averaged over several
observations taken over relatively short intervals. Occasionally, the observer
would recompute the averages in a subsequent publication based on a different
combination of the measurements. This sometimes led to redundant listings for
the same measurements. We cross-compared the S51 tabulation and WDS catalog with
the original publications, and adopted the values published most recently by the
observers. Additionally, in compiling his data, S51 sometimes averaged
observations that had not been averaged by the original observers, and that are
now listed individually in the WDS\null. In these instances, we adopted the
individual measurements as listed in the WDS\null. Table~6 indicates the
observations that we removed from the S51 and WDS listings, and which
measurements we used to replace them.

Table 7 gives the complete list of edited ground-based measurements that were
initially used in our orbit fit. Some of the observations were badly discrepant
and were removed in our final fit; these are identified in the table by a
superscript c in the first column. Our fitting procedure and rejection process
are described below in \S4.2. In his tabulation, S51 had corrected the
position angles for precession to the J2000 equinox; in our Table~7 we have
similarly corrected the position angles for the ground-based measurements after
1932 to the J2000 equinox (except for CCD and adaptive-optics observations,
which we assumed to be reported for J2000).


%

\subsection{Elements of the Relative Visual Orbit}

We fitted a visual orbit simultaneously to the \HST\/ and ground-based  
measurements (Tables~3 and 7 respectively; for the \HST\/ WFC3 data, we used the
``F953N average'' values). We used a Newton-Raphson method to minimize $\chi^2$
by calculating a first-order Taylor expansion for the equations of orbital
motion. For the \HST\/ data we used the measurement errors directly from Table~3
in computing $\chi^2$. The ground-based observers typically did not estimate
errors for their measurements, so we adopted an iterative approach to optimize
the weighting of the ground-based data in our orbit fit, and to reject outliers.
In the first step of the iterative procedure, we fit an orbit to the
ground-based data only and applied uniform uncertainties to these measurements
to force the reduced $\chi^2_\nu$ to equal unity (where $\nu$ is the degree of
freedom). In the second step, we used these scaled uncertainties to fit an orbit
simultaneously to the ground-based and \HST\/ measurements.  We used a
sigma-clipping algorithm to reject any ground-based data point whose residual
was more than three times the standard deviation of the residuals for the full
data set. We repeated this procedure until no additional data points were
rejected. The final data set contained the 57 measurements listed in Tables~3
and 7 (18 from \HST\/ and 39 ground-based retained in the solution). The adopted
uncertainties for the ground-based separations were $\pm$0\farcs187; we
propagated this value to the position angle by assuming equal uncertainties in
the right-ascension and declination directions. The historical measurements
removed from the fit through sigma clipping are flagged in Table~7. 
Many of the rejected observations were made between 1914 and 1929, when the
visual measurements were extremely difficult---or even, as suggested by S51, of
doubtful reality.

Table~8 lists the final parameters for the visual orbit. The uncertainties were
computed from the diagonal elements of the covariance matrix. We also
investigated a solution using {\it only\/} the \HST\/ measurements; this
solution produced uncertainties averaging about 60--70\% larger than those
presented in Table~8, with the error in the orbital period more than doubled. An
additional, and probably final, \HST\/ observation will be scheduled in 2016,
but we expect that the historical ground-based data will continue to be an
essential part of the best orbital solution. 

In Figure~4 we plot the data points, both \HST\/ and ground-based, and the
orbital fit. The positions of Procyon~B that are predicted from our orbital
elements in Table~8 are marked with open blue circles, for the \HST\/
observations only. At the scale of Figure~4, the observed \HST\/ data (filled
black circles) are so precise that they appear to lie exactly at the centers of
the open blue circles. For a better visualization of the errors, the two panels
of Figure~5 show the residuals of the \HST\/ observations from the positions
predicted by our orbital elements, in right ascension and declination. The units
are now milliarcseconds (mas), rather than the arcseconds of Figure~4. The error
bars are those given in Table~3, converted from separation and position angle to
right ascension and declination. Based on the residual plots, there is no
evidence within those errors for perturbations of the orbit by a third body. (We
return to this point in~\S6.) 


\section{Determining Dynamical Masses}

\subsection{Parallax and Semimajor Axis of Procyon A}

In addition to the elements of the relative orbit listed in Table~8, we need two
further quantities in order to determine dynamical masses for both stars: the
absolute parallax of the system, and the semimajor axis of the absolute motion
of Procyon~A on the sky.

There are three recent independent determinations of the parallax: (1)~G00
obtained it from measurements of a series of $\sim$50 plates taken at the USNO
1.55-m reflector between 1985 and 1990; (2)~the parallax was measured by \Hipp\/
(we use the value from the new reduction by van Leeuwen 2007); and (3)~it was
measured with the Allegheny MAP by Gatewood \& Han (2006). These results are in
good agreement, and we adopt a weighted mean of the parallax values, as given in
the top part of Table~9. 

%


G00 determined the semimajor axis of Procyon~A's motion from $\sim$600 exposures
on plates obtained at six different observatories, from 1912 to 1995. We have
adopted their result, $a_A=1\farcs232\pm0\farcs008$, as given in the bottom of
Table~9. It agrees fairly well with a value of $1\farcs217\pm0\farcs003$
obtained by S51 from a subset of the same photographic
material.\footnote{Strand actually gave a probable error of $\pm\!0\farcs002$,
which we have converted to standard error here.} The G00 result differs by a
larger amount from the $1\farcs179\pm0\farcs011$ found by I92 from a combined
analysis of RVs and a re-analysis of S51's astrometry. 

Our decision to adopt the G00 value of $a_A$ over that of S51---despite his
quoted uncertainty being smaller---is based on the significant advantages of the
G00 study.  These include the use of plates spanning twice the time baseline
($\sim$80~yr vs.\ $\sim$40~yr); digital centering with a laser-encoded PDS
microdensitometer as opposed to visual centering with a single-screw measuring
engine; and computer-calculated plate transformations using scores of reference
stars compared to the four reference stars used in Strand's manual
calculations.  It is possible that this last limitation of having to rely on
just four reference stars might have caused Strand to underestimate the
uncertainty in $a_A$. As a check on the uncertainty estimated by G00, Elliott
Horch kindly reprocessed the 593 measures from G00 using his independent
orbit-element code. The uncertainty in $a_A$ was confirmed to be
$\pm\!0\farcs008$.

There is also the question of the sensitivity of the value of $a_A$ derived by
G00 to the adopted orbital elements, given our new and more precise
determination of those elements. We investigated this by assuming the values
in our Table~8 for all elements except the semimajor axis, and then
reprocessing the photographic measures of G00, solving only for $a_A$\null.  The
result is unchanged and robust, with values of $a_A$ ranging from $1\farcs231$
to $1\farcs234$, depending on the degree of ``outlier'' trimming. For these
reasons, we have adopted the values of $a_A$ and its uncertainty as given by
G00.

\subsection{Comparison with Radial Velocities}

We did not use RV information in our orbital solution, which was based purely on
astrometric data. The RVs, however, provide a useful check on the validity of
our final results.

Our orbital elements, along with the parallax and the semimajor axis of
Procyon~A's motion, allow us to predict the RV of Procyon~A, apart from a
constant offset due to the center-of-mass motion of the binary system. In
Figure~6 (top), we compare our predictions with absolute RVs published by I92
(from photographic spectrograms, 1909--1985), and a single absolute RV by Mosser
et al.\ (2008, from RV speedometer). Also plotted are relative RVs by Innis et
al.\ (1994; from RV spectrometer data, 1986--1990). Innis et al.\ adjusted their
velocity zero-point so that their RVs would match the I92 orbit predictions in
the mean. We have applied I92's gamma-velocity of $-4.115\,\kms$ to our
predicted RVs. 

I92 also published a separate set of precise RVs of Procyon~A, measured using a
hydrogen-fluoride absorption cell, obtained over the interval 1980--1991. These
velocities are on a relative scale. In Figure~6 (bottom), we compare the RVs
predicted by our orbital elements with these velocities; we have arbitrarily
shifted the zero-point of our predictions to match the measurements in the mean.
Both of these figures show that our parameters of the Procyon system are able to
predict the RV measurements very well.

%
%
%

\subsection{Dynamical Masses}

Table~10 lists the dynamical masses that result from our adopted parameters. We
used the usual formulae for the total system mass, $M = M_A+M_B = a^3/(\pi^3 \,
P^2)$, and for the individual masses, $M_A = M\,(1-a_A/a)$ and $M_B =
M\,a_A/a\,$; in these equations the masses are in $M_\odot$, $a$ and $\pi$ in
arcseconds, and $P$ in years.

In Table~11 we present the error budgets for the masses of Procyon~A and B,
based on the adopted random uncertainties of each of the parameters.\footnote{A
potential source of systematic uncertainty is errors in the plate scales of the
\HST\/ cameras. Gonzaga \& Biretta (2010) state a fractional uncertainty of
$\pm$0.0003 for the WFPC2 plate scale, and for the WFC3 plate scale we derived a
similar fractional uncertainty of $\pm$0.00018. These imply a systematic
uncertainty of about $0\farcs0013$ for the semimajor axis, $a$. Table~11 shows
that a systematic error of this magnitude contributes negligibly to the random
errors in the dynamical masses.} For Procyon~A, the mass error is dominated
almost entirely by the uncertainty in the parallax. For the WD companion,
Procyon~B, the parallax is again responsible for the majority of the error, but
the uncertainty in the semimajor axis of Procyon~A's astrometric motion also
contributes significantly. Unfortunately, the mass uncertainties are unlikely to
be reduced in the near future, because Procyon is too bright for its parallax to
be measured by the {\it Gaia\/} mission (D.~Pourbaix, private
communication).\footnote{{\it Gaia\/} may provide a slight improvement in the
correction of ground-based parallaxes to absolute, by determining parallaxes for
the reference stars used in the ground-based determinations.}

\section{Limits on Third Body}

%
%
%

%
%
%
%
%
%
%
%
%
%
%
%
%
%

As discussed in \S4.2 and shown in the residuals plotted in Figure~5, we
detected no significant perturbations in our orbital fit to the \HST\/
astrometry. These results allow us to place limits on the presence of third
bodies orbiting either star in the Procyon system. 

The stability of planets orbiting the individual stars in a binary system has
been studied numerically by, among others, Holman \& Wiegert (1999). Using the
results in their Table~3, and the parameters of the present-day binary, we find
that the longest periods for stable planetary orbits in the Procyon system are
about 3.7~yr for a planet orbiting Procyon~A, and 2.8~yr for one orbiting
Procyon~B\null.

We calculated the semimajor axes of the astrometric perturbations of both stars
that would result from being orbited by planetary companions of masses ranging
from 5 to $25\,\Mjup$ (where $\Mjup$ is the mass of Jupiter,
$0.000955\,M_\odot$), and for orbital periods up to the stability limits given
above. The results are plotted in Figure~7. Based conservatively on Figure~5, a
periodic astrometric perturbation of either star with a semiamplitude larger
than $\sim\!3$~mas would have been detected. The data in Figure~7 then indicate
that a companion of Procyon~A of $\sim\!5\,\Mjup$ or less could escape
astrometric detection. At $\sim\!10\,\Mjup$, only an orbital period longer than
$\sim$1.5~yr would have led to detection in our data. Progressively more massive
planets orbiting Procyon~A would have been detected more easily, except at the
shortest orbital periods. Thus, in general, our limits are not competitive with
what can be achieved with high-precision RV data (apart from orbits viewed at
very low inclinations).

Our limits are more useful for Procyon~B, for which a precision RV study is
impractical. A $\sim\!5\,\Mjup$ companion with a period longer than $\sim$2~yr
is excluded, and for $\sim\!10\,\Mjup$ the lower-limit period is $\sim$0.5~yr.

\section{Astrophysics of Procyon A}

We now turn to discussions of the astrophysical implications of our
dynamical-mass results for both stellar components of Procyon. We start in this
section with the primary star, Procyon~A, and then discuss the WD Procyon~B in
\S8.

\subsection{Asteroseismology}

With the advent of asteroseismology, Procyon~A was recognized as a unique object
for exploring non-radial stellar oscillations. Oscillation frequencies are
particularly sensitive to boundaries between radiative and convective regions.
Stars near the main sequence in the mass range near that of Procyon~A are
believed to exhibit a convective core, a radiative envelope, and a very thin
outer convection zone (e.g., Guenther \& Demarque 1993).

Helioseismology shows that diffusion of helium and heavy elements in the solar
interior significantly affects the solar oscillation frequencies, and it is
expected also to play a role in the radiative envelope of Procyon~A\null. In
addition, the convective core overshoot at the core's edge must be taken into
account. The amount of core overshoot in a star of this mass is not well known.
It strongly affects the morphology of the evolutionary track and the
evolutionary rate in the post-main-sequence phase of evolution where Procyon
lies.

The oscillation spectrum of Procyon~A has been obtained from ground-based
radial-velocity measurements (Arentoft et al.\ 2008), and from intensity
observations with the space mission {\it MOST\/} (see Guenther et al.\ 2008). A
Bayesian statistical study of the asteroseismic data, based on a large grid of
stellar-evolution tracks, was carried out by Guenther, Demarque, \& Gruberbauer
(2014; hereafter GDG14). Their tracks spanned the mass range 1.41 to
$1.55\,M_\odot$\null. Other grid parameters (see Table~2 of GDG14) covered the
following variables: (1)~the helium and heavy-element contents by mass ranged
from $Y=0.26$ to 0.31 and $Z=0.014$ to 0.031; (2)~the mixing-length parameter,
$\alpha$, in the thin outer convection zone ranged from 1.7 to 2.5; and (3)~the
core-overshoot parameter, $\beta$, initially ranged from 0.0 to 1.0 times the
local pressure scale height, $H_p$. Because of unanticipated evidence for large
convective overshoot, the grid was eventually extended to $\beta$ values as
large as $2.0\,H_p$. Three quantities were selected as priors in the
calculations, namely the mass of Procyon~A (from G00), and its position ($\log
L/L_\odot$ and $\log\Teff$) in the HR~diagram (HRD; see Table~1 of GDG14).

The strongest result of the GDG14 analysis was that all of the most probable
theoretical models (with or without core overshoot, with adiabatic or
non-adiabatic model frequencies, with or without diffusion in the radiative
envelope, and including or not including priors for the observed HRD location
and mass) were found to have masses within 1$\sigma$ of the mass inferred from
observations of $1.497 \pm 0.037\, M_\odot$, as published by G00. The error bar
for the most probable theoretical mass is still large, and more precise
oscillation frequencies will be needed in the future to reduce it. But it is
encouraging that this result is in such good agreement with the dynamical mass
of $1.478 \pm 0.012\,M_\odot$ derived from our observational analysis in the
present paper.


 Another result of the Bayesian analysis is relevant. The most probable
models were characterized by substantial overmixing beyond the formal boundary
of the convective core, with values of $\beta$ as high as $1.0\,H_p$ or even
larger. This result exceeds the expected value of $\beta = 0.2$ or less,
generally accepted for core convective overshoot in similar stars. This may be
evidence for diffusive mixing beyond the standard overshoot region, as recently
discussed by Moravveji et al.\ (2015) in the case of the more massive star
KIC~10526294.  A full understanding of this result will require continued
seismic monitoring of Procyon~A to improve the precision of the oscillation
frequencies, as well as more sophisticated modeling in the overshoot region.



\subsection{The Age of Procyon A}

Knowing the age of Procyon~A is critical to understanding the past evolution of
the binary system. In conjunction with the cooling age of the companion WD, the
age of Procyon~A allows us to estimate the original mass of the Procyon~B
progenitor (see L13 and the discussion in \S8.2 below).


We constructed grids of stellar evolutionary tracks for stars with the dynamical
mass of $1.478\,M_\odot$ derived in the present paper, following them from the
zero-age main sequence (ZAMS) to the subgiant branch.  The tracks were 
calculated under the assumption of single-star evolution, i.e., no interaction
between Procyon~A and its companion during the course of its evolution from the
ZAMS to the present.

We calculated models that either include or ignore the effects of element
diffusion, and for amounts of convective-overshoot efficiency at the edge of the
convective core of $\beta=0$, 0.2, and $1.0\,H_p$. 
A standard model of convective core overshoot was adopted, as described in
GDG14.  The temperature gradient in the overshoot region was constrained to be
the local radiative temperature gradient. This situation has been described as 
``overmixing,'' as opposed to ``penetrative convection,'' where the temperature
gradient is adiabatic (see Zahn 1991; GDG14).
We used a near-solar metallicity of $Z=0.02$, and adjusted the hydrogen
abundance in the ZAMS starting model so as to ensure that each track passed
through the Procyon~A error box (from Do{\u g}an et al.\ 2010) in the HRD. 

Figure~8 displays three of these tracks, constructed with the stellar-evolution
code YREC (Demarque et al.\ 2008). All of these models include helium and
heavy-element diffusion in the radiative envelope, using the formalism of
Bahcall \& Loeb (1990).  Core overshoot is the major uncertainty in determining
the ages of these models. The track plotted in red assumes no core overshoot
($\beta=0$); the track in green has a standard value of core overshoot
($\beta=0.2$); and the track in blue has large core overshoot ($\beta=1.0$). The
hydrogen contents in the ZAMS starting models were $X= 0.672$ (red track), 0.680
(green track), and 0.716 (blue track). These abundances are all consistent with
accepted uncertainties in $X$.

The red and green tracks are morphologically similar to tracks used in previous
studies of Procyon~A\null. In particular, note Procyon~A's position just below
the well-known leftward ``hook,'' due to core hydrogen exhaustion. This near
coincidence was, until the advent of precision asteroseismology, a major source
of ambiguity in identifying the precise evolutionary status of Procyon~A\null.
However, seismology clearly placed Procyon~A in the core-burning phase of 
evolution; but as discussed above it also surprisingly revealed the presence of
extensive mixing in the interior outside the convective core (see GDG14). Due to
the very large amount of overshoot, the blue track has a quite different
morphology from the other two, and it also evolves more slowly. In this case,
Procyon~A lies well before the hydrogen-exhaustion phase. 

The Procyon~A ages based on the red, green, and blue tracks are 1.673, 1.817,
and 2.703~Gyr. If one accepts the main results from the GDG14 Bayesian
statistical analysis of the seismic data, then the preferred age is close to
2.70~Gyr.  
Such an age may be an upper limit. While a minimum age near 1.8~Gyr (as found by
both L13 and GDG14) seems well established, there remains an uncertainty in the
maximum age, which depends sensitively not only on the amount of chemical mixing
from the core but also on the composition profile and structure above the core
edge in the envelope. The recently published seismological study by Moravveji et
al.\ (2015) of KIC~10526294, a $3\,M_{\odot}$ star near the main sequence, shows
that the frequencies observed by the {\it Kepler\/} mission can be tightly
fitted to a diffusion model in the overshoot region. Improved oscillation
frequencies of the same quality will be needed to produce a similar result for
Procyon~A\null. Finally, it should also be emphasized 
that the validity of this discussion rests upon the assumption
of single-star evolution at constant mass for the A component (see below).

\section{Astrophysics of Procyon B}

\subsection{Testing White-Dwarf Physics}

Procyon~B lies mostly hidden in the glare of its much brighter primary star.
Virtually all that was known about it in the pre-\HST\/ era was based only on
its astrometric properties and approximate brightness estimates. However, using
\HST, Provencal et al.\ (1997) acquired WFPC2 images of Procyon~B in several
wide- and narrow-band filters, with effective wavelengths ranging from 1600 to
7828~\AA, and covering most of the star's spectral energy distribution
(SED)\null. From these photometric data they deduced a helium-composition
photosphere, and estimated the star's effective temperature ($\Teff = 8688 \pm
200$~K) and radius ($R_B=0.0096 \pm 0.0005\,R_\odot$). Based on the I92
astrometric mass of $0.622 \pm 0.023\, M_\odot$, this relatively small radius
called into question the assumption of the CO degenerate core that would be
expected for a WD of this mass. Provencal et al.\ instead suggested the
remarkable possibility of an iron core---placing the star in an ``iron
box''---as implied by the zero-temperature WD mass-radius relations of Hamada \&
Salpeter (1961).  

The nature of Procyon B became clearer five years later when P02 used \HST\/ to
obtain a series of Space Telescope Imaging Spectrograph (STIS) spectra, covering
1800 to 10,000~\AA\null. These revealed the presence of C$_2$ Swan bands, as
well as absorption features due to \ion{C}{1}, \ion{Mg}{2}, \ion{Ca}{2}, and
\ion{Fe}{1}\null. Balmer lines are absent. Along with the earlier results, these
features show the star to be a DQZ WD, i.e., having a He-dominated atmosphere,
but also containing carbon (Q) and heavier metals (Z)\null. Model-atmosphere
fitting to the STIS spectra resulted in a significantly lower $\Teff$ of $7740
\pm 50$~K, and a correspondingly larger radius of $0.01234 \pm 0.00032\,
R_\odot$, based on a $V$ magnitude of $10.82 \pm 0.03$ obtained from the
observed SED\null.  This radius, along with a lowered astrometric mass of $0.602
\pm 0.015\, M_\odot$ from G00, removed the earlier discrepancy with the
mass-radius relation for CO-core WDs.  

Our new dynamical mass for Procyon B allows refinement of a number of its
astrophysical parameters and a stringent test of theoretical WD models. We have
slightly modified the radius determined by P02, by adjusting for our adopted
parallax, obtaining $R_B = 0.01232 \pm 0.00032\,R_\odot$\null.  In Figure~9, we
compare theoretical predictions with our new parameters for Procyon~B\null. We
use theoretical modeling data from the Montreal photometric tables\footnote{{\tt
http://www.astro.umontreal.ca/$^\sim$bergeron/CoolingModels}. These tables are
based on evolutionary sequences and model atmospheres calculated by Holberg \&
Bergeron (2006), Kowalski \& Saumon (2006), Tremblay et al.\ (2011), and 
Bergeron et al.\ (2011).} for WDs with pure-helium atmospheres and CO cores. The
top panel in Figure~9 shows the location of Procyon~B in the theoretical HRD
($\log L/L_\odot$ vs.\ $\log\Teff$), along with the model cooling tracks for DB
WDs with masses of 0.5, 0.6, and $0.7\,M_\odot$\null. The location of Procyon~B
in the HRD is in excellent agreement with that expected for a WD of our
dynamical mass of $0.592 \pm 0.006 \, M_\odot$\null.  Also shown in the top
panel of Figure~9 are isochrones for ages of 1, 1.25, and 1.5~Gyr, again based
on the Montreal tables. By interpolation in the theoretical data, we estimate
the cooling age of Procyon~B to be $1.37 \pm 0.04$~Gyr. 

In the bottom panel of Figure~9, we plot the position of Procyon~B in the
mass-radius plane. It is compared with a theoretical mass-radius relation for a
He-atmosphere CO-core WD with $\Teff = 7740$~K, obtained through
interpolation in the Montreal tables. The observed mass and radius are in
excellent agreement with the theoretical relation. Also plotted is the Hamada \&
Salpeter (1961) mass-radius relation for zero-temperature WDs composed of
$^{56}$Fe, which was consistent with the parameters of Procyon~B given by
Provencal et al.\ (1997); with our revised parameters, there is no longer
agreement with Fe---as first shown by P02. 

The surface gravity of Procyon~B (in cgs units), based on the mass and radius,
is $\log g = 8.028 \pm 0.023$. Unfortunately, without Balmer lines, there are no
gravity-sensitive features in the \HST\/ spectra that would test for consistency
with this value. The predicted gravitational redshift is $30.46 \pm 0.85\,\kms$,
but Procyon~B possesses no detectable H$\alpha$ line from which the redshift
could be measured using traditional techniques.  Onofrio \& Wegner (2014) have
recently attempted to measure wavelength shifts in the archival \HST\/ spectra
of Procyon~B, using features of \ion{Ca}{2}, \ion{Mg}{2}, and C$_2$. They appear
to have detected the gravitational redshift, but uncertain corrections for
pressure shifts are needed.

\subsection{Procyon B Progenitor and the Initial-to-Final-Mass Relation}

L13 made a comprehensive analysis of the existing data on Procyon~A and B,
aiming to determine a consistent picture of the system's evolution. For
Procyon~B, L13 adopted the P02 effective temperature, but assumed a mass of
$0.553 \pm 0.015\,M_\odot$, a value $\sim\!0.04\,M_\odot$ lower than the
dynamical mass determined in the present paper.\footnote{The lower mass used by
L13 was from a 2012 private communication from G.H.S. and H.E.B.; at that time
we still had not done the rigorous astrometry of our \HST\/ images described in
the present paper, and also were using the smaller value of $a_A$ from I92.}
From these parameters, L13 obtained a cooling age of 1.19~Gyr for
Procyon~B\null. Combined with the age of Procyon~A, which they determined to be
1.87~Gyr from its position in the HRD, this implied a main-sequence lifetime of
only 0.68~Gyr for the progenitor of the WD, corresponding to an initial mass of
$2.59\,M_\odot$. As L13 noted (their Fig.~2 and associated text and references),
these results placed Procyon~B significantly below the initial-to-final-mass
relation (IFMR) established from studies of WDs in open clusters. A
$2.59\,M_\odot$ progenitor would be expected to produce a WD with a mass of
about $0.69\,M_\odot$ (cf.\ Ferrario et al.\ 2005).



%
%

As discussed in \S7.2, the age of Procyon~A may be considerably greater than
adopted by L13. This is the case if we use the evolutionary track with large
core overshoot, as favored by the GDG14 seismologic analysis. For a Procyon~A
age of 2.70~Gyr, and our cooling age of 1.37~Gyr for the WD, the main-sequence
lifetime of the progenitor of Procyon~B was 1.33~Gyr. This corresponds to a ZAMS
mass of about $2.2\,M_\odot$ (if the progenitor had a large core overshoot of
$\beta=1.0$ like Procyon~A), or a lower initial mass of about $1.9\,M_\odot$ (if
it had a ``normal'' overshoot of $\beta=0.2$). The Ferrario et al.\ (2005) IFMR
predicts WD masses of 0.65 or $0.625\,M_\odot$ for such initial masses. Thus
there remains a discrepancy, albeit smaller than found by L13, and within the
cosmic scatter in the relation (e.g., Fig.~2 in L13).

If future observations force an unlikely revision in the current  interpretation
of the seismic data, Procyon~A's age could in principle be as low as
$\sim$1.8~Gyr, resulting in an initial mass possibly as high as about
$3\,M_\odot$ for Procyon~B---and a much more severe disagreement with the mean
IFMR\null. 

\subsection{Atmospheric Carbon and Heavy Elements}

Procyon B presents an unusual case of a WD with the rare DQZ spectral type being
a companion of a main-sequence (or slightly evolved) star. A somewhat similar
system, HR~637 (GJ~86), was recently studied by Farihi et al.\ (2013). The K0~V
primary in this binary is orbited in a 15.9-day period by a Jovian planet of
perhaps 4.4--$4.7\, \Mjup$, detected through RV measurements (Queloz et al.\
2000). The K dwarf also has a WD companion in a more distant orbit, with a
period estimated at several hundred years. Farihi et al.\ used \HST/STIS to
obtain spectra of the resolved WD companion. They found it to be He-rich, with
C$_2$ absorption bands (spectral type DQ6), and having a remarkably similar
temperature, mass, and atmospheric carbon content to those of Procyon~B\null.
The carbon in cool DQ atmospheres is likely intrinsic to the star. However, the
heavier metals (e.g., Ca, Mg, and Fe) seen in Procyon~B, but not in HR~637~B,
are probably accreted from an external source.

The source of heavy elements accreting onto the photospheres of single DA and DB
WDs is usually considered to be a circumstellar debris disk, composed of rocky
material (e.g., Jura 2003). Such disks likely form when the WD tidally disrupts
terrestrial planets, asteroids, or planetesimals. Although no heavy elements
were observed in HR~637~B, Farihi et al.\ discussed their presence in
Procyon~B\null. They examined the heavy-element accretion rates necessary to
account for the abundances of Ca and Mg in the Procyon~B photosphere, along with
the lack of hydrogen, and ruled out accretion from the interstellar medium or a
stellar wind from Procyon~A\null. They instead argued for a circumstellar disk
around Procyon~B as the heavy-element source. 

Such a debris disk would be very compact, within a few tenths of a solar radius,
and not significantly influenced by the gravity of Procyon~A\null. The object(s)
that formed the disk were unlikely to have originated in a protoplanetary disk
around the Procyon~B precursor, since planet formation would have been confined
to within $\sim$2.3~AU (Holman \& Wiegert 1999), and any such planetesimals
would likely have been destroyed during the red-giant phases. Instead, Farihi et
al.\ argue that the reservoir could be a much larger disk enclosing the entire
binary. They also conclude that it was unlikely that any Jovian-mass objects
ever formed around either Procyon~A or B, since their formation would have been
confined to within the snow limits of each star. This is certainly consistent
with the lack of any dynamical evidence for existence of $\sim$5 to $10\,\Mjup$
third bodies, as reported in our \S6.

Another alternative is that the polluting material results from
``second-generation" planets, as described by Perets \& Kenyon (2013). They
suggest that a portion of the wind from an AGB star may be captured by a binary
companion (in this case, Procyon~A), creating a disk in which planets may form.
However, in order to pollute the WD atmosphere, an asteroid or planetesimal born
in this disk would then have to be ejected from its orbit around A and into the
vicinity of Procyon~B.

\section{Past Evolution of the Procyon System}

The discussion in the previous two sections treated both components of the
Procyon system as if they have evolved as single stars. Such an analysis does
lead to a reasonably consistent picture, with a primary star whose position in
the HRD can be reproduced with theoretical tracks based on the star's observed
mass (although with indications of unusually efficient core overshoot), and a
reasonably well-behaved WD companion (although with hints that its mass is
somewhat lower than expected).

The periastron separation of A and B in the present-day orbit is 9.1~AU\null. If
the progenitor of B had a mass of $\sim\!2.2\,M_\odot$, as deduced in the
previous section, then the total mass of the system was reduced from
$\sim\!3.7\,M_\odot$ to its present value of $2.07\,M_\odot$ due to 
evolutionary mass loss from the progenitor. Under the assumption that the mass
loss was on a timescale slow compared to the orbital period (cf.\ Burleigh et
al.\ 2002, \S2), this implies that the periastron separation of the pair was
only $\sim$5.1~AU in the progenitor system. 


At such a minimum separation, the progenitor was likely to have avoided
Roche-lobe overflow. This is consistent with the high orbital eccentricity 
(0.40) in the present system, which appears to rule out a phase in which the two
stars shared a common envelope (during the giant or AGB phase of the initial B
component), because it would have led to rapid circularization of the orbit---if
not a spiralling down to a shorter period or even a merger. The high
eccentricity thus sets indirectly an upper limit on the initial mass of
Procyon~B\null. However, the eccentricity may have favored strong periodic tidal
interaction at times of closest approach between the two stars. The unusually
large mixing detected by seismology could then be the result of such tidal
interaction.

A more extreme interaction may have occurred during wind mass loss from the B
progenitor when it was a red giant or AGB star, and during ejection of a
planetary nebula. This may have led to mass transfer from the WD progenitor onto
Procyon~A (see Wegner 1973; Fuhrmann et al.\ 2014)---in addition to the
hypothetical disk and second-generation planet formation around A discussed in
the preceding section. Thus the original Procyon~A could have been less massive
than at present, and therefore more slowly evolving. The main effect of
accretion would have been to speed up the rate of evolution of Procyon~A from
that of a lower-mass star to its rate at the present time.

\section{Summary }

Based on our analysis of two decades of precise astrometry of the Procyon system
with the {\it Hubble Space Telescope}, combined with historical measurements
dating back to the 19th century, we have derived dynamical masses for both
components. The F5 subgiant Procyon~A is found to have a mass of $1.478 \pm
0.012 \, M_\odot$, and the Procyon~B white-dwarf companion has a mass of $0.592
\pm 0.006 \, M_\odot$.  We find no evidence for perturbations due to third
bodies in the system, at levels down to about 5--$10\,M_{\rm Jup}$\null. 

The mass of Procyon~A is in excellent agreement with theoretical predictions
based on asteroseismology and its position in the H-R diagram. However, a
surprisingly high amount of core convective overshoot, compared with that
usually adopted for individual stars and stars in open star clusters, is
required to achieve this agreement. If correct, this implies that the age of
Procyon~A is about 2.7~Gyr.

The position of Procyon~B in the H-R diagram is in excellent agreement with a
theoretical cooling track for a white dwarf of its measured mass, and implies a
cooling age of 1.37~Gyr. In the mass-radius plane, Procyon~B's location is in
agreement with theoretical predictions for a carbon-oxygen white dwarf with a
helium-dominated atmosphere.  The mass of its progenitor, if the age of A is
2.7~Gyr, was about $1.9\,M_\odot$ if the progenitor had a ``normal'' amount of
core overshoot, or about $2.2\,M_\odot$ if it had a larger amount similar to
that of A\null. In either case, the mass of the white dwarf is lower than
expected based on the mean initial-to-final-mass relation for single stars in
open clusters, although still within the cosmic scatter.

Although treating both stars as if they have evolved separately leads to a
fairly consistent interpretation of the system, we point out that in the
progenitor system the two stars were actually relatively close to each other
($\approx$5~AU) at every periastron passage. Thus the stars may have been
affected by tidal interactions and even mass capture from a red-giant wind, and
their actual evolutionary histories may have been much more complicated than the
simple picture presented here.

\acknowledgments

Support was provided by NASA through grants from the Space Telescope Science
Institute, which is operated by the Association of Universities for Research in
Astronomy, Inc., under NASA contract NAS5-26555. J.B.H. acknowledges support
from NSF grant AST-1413537. We are grateful to Elliott Horch for help in
verifying the orbit fit to the photographic data, and to Peter Eggleton, David
Guenther, and Bill van Altena for several helpful discussions.

{\it Facilities:} \facility{\HST\/ (WFPC2, WFC3)}


\clearpage


\begin{figure}
\begin{center}
\includegraphics[width=5.5in]{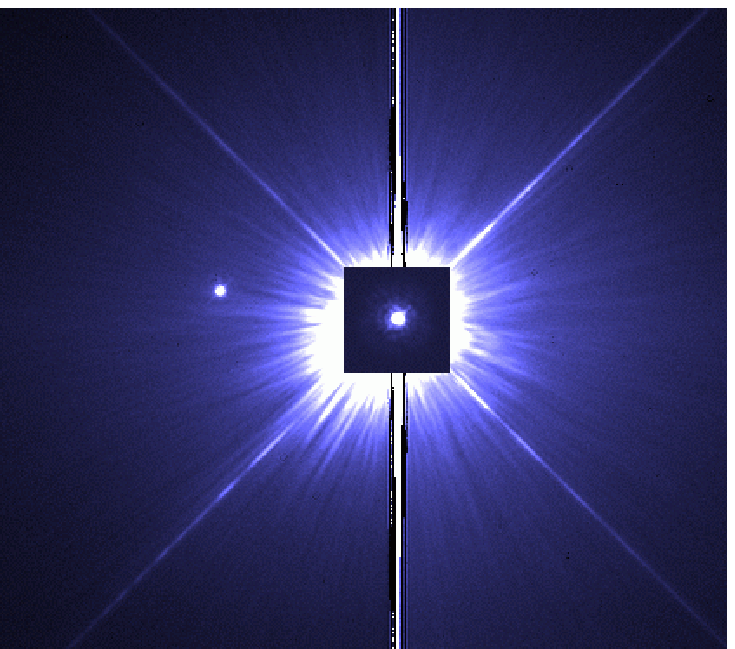}
\figcaption{False-color \HST\/ WFPC2 image of Procyon, from frames obtained in
the near-ultraviolet F218W filter on 1997 November~27. An inset showing the
unsaturated Procyon~A from a 0.14-s exposure is superposed on a 100-s exposure,
taken at the same telescope pointing. The white dwarf Procyon~B is easily
resolved, at a separation of $4\farcs706$. In the near ultraviolet the measured
brightness difference is 8.5~mag. 
}
\end{center}
\end{figure}

\clearpage


\begin{figure}
\begin{center}
\includegraphics[width=5.5in]{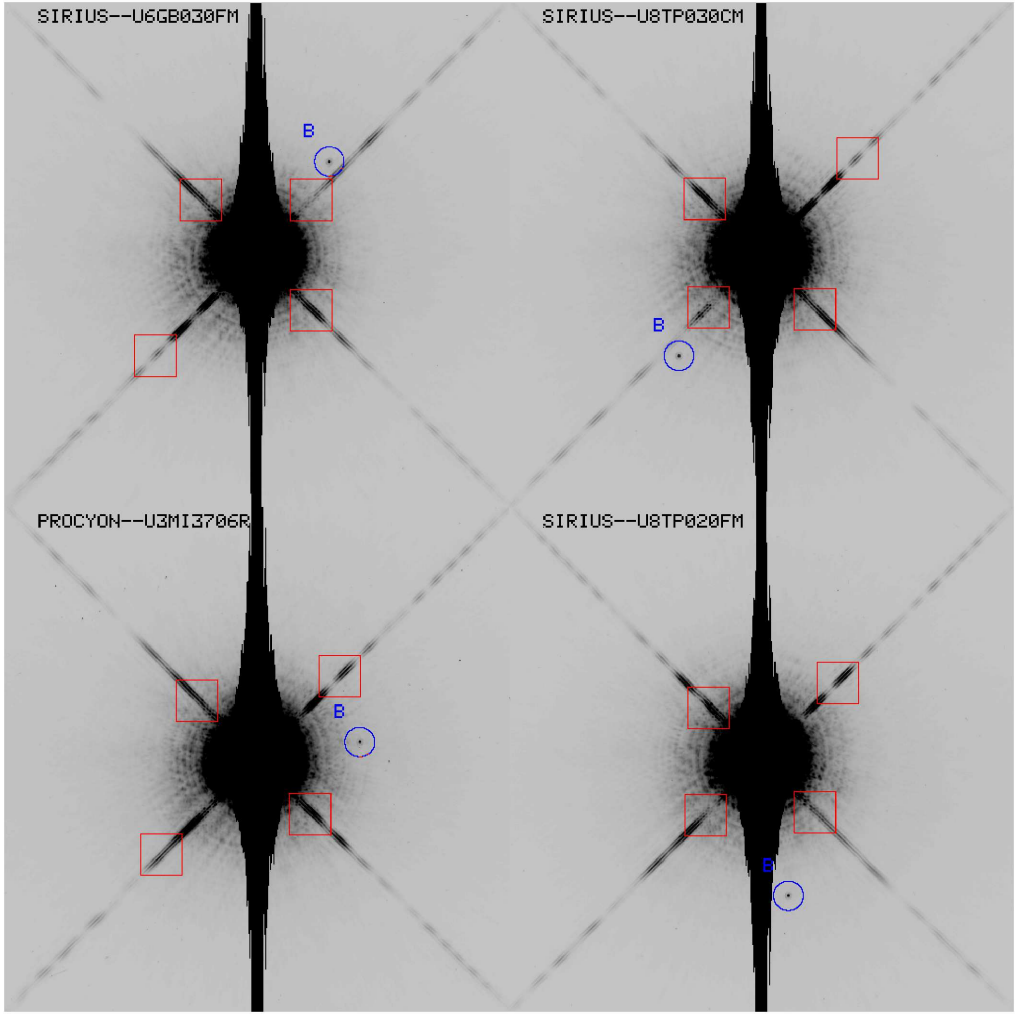}
\end{center}
\caption{Each of these four subpanels shows a $512\times512$ pixel region from
WFPC2 images, centered on either Procyon A or Sirius A, with specific image
numbers indicated within the figure.  These $23''\times23''$ frames show the
strongly oversaturated A at center, with the much fainter B component circled
and labeled.  The red boxes indicate the regions within each diffraction spike
that were used to determine the centroid location of A; see text for details. }
\label{fig:panF1042M}
\end{figure}


\begin{figure}
\begin{center}
\includegraphics[width=5.5in]{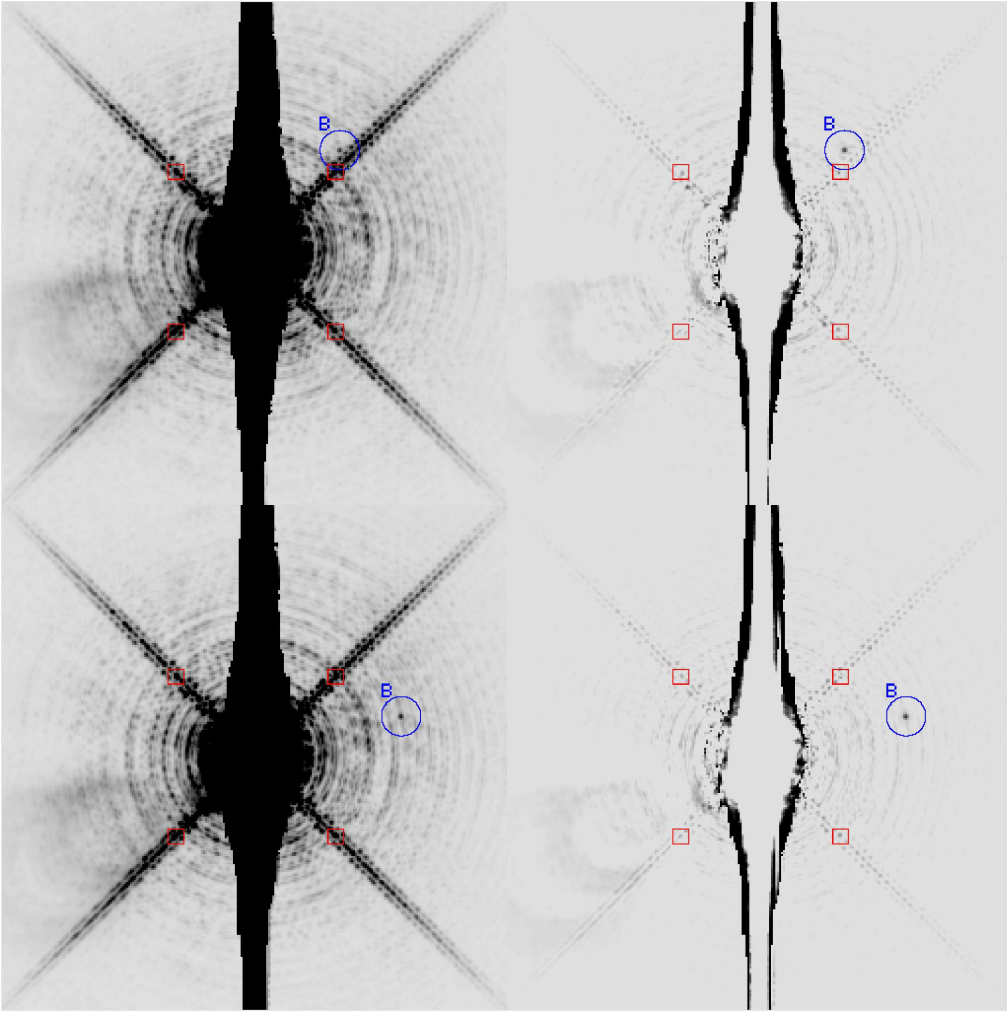}
\end{center}
\caption{
The upper two images show $256\times256$ pixel ($10\farcs1\times10\farcs1$)
regions centered on Procyon A from a WFC3 frame ({\tt ibk701010}) obtained on
2011 Feb 7, and the lower two images show Procyon A from another frame ({\tt
ibti01010}) taken on 2012 Mar 9. The two left-hand panels are the direct images,
while the right-hand panels are after a best-fit representation of the
over-sampled saturated PSF has been subtracted. The red squares indicate the
regions on the diffraction spikes that were used to fix the PSF centroid, while
the location of Procyon~B is circled and labeled.
}
\end{figure}

\begin{figure}
\begin{center}
\includegraphics[width=6.5in]{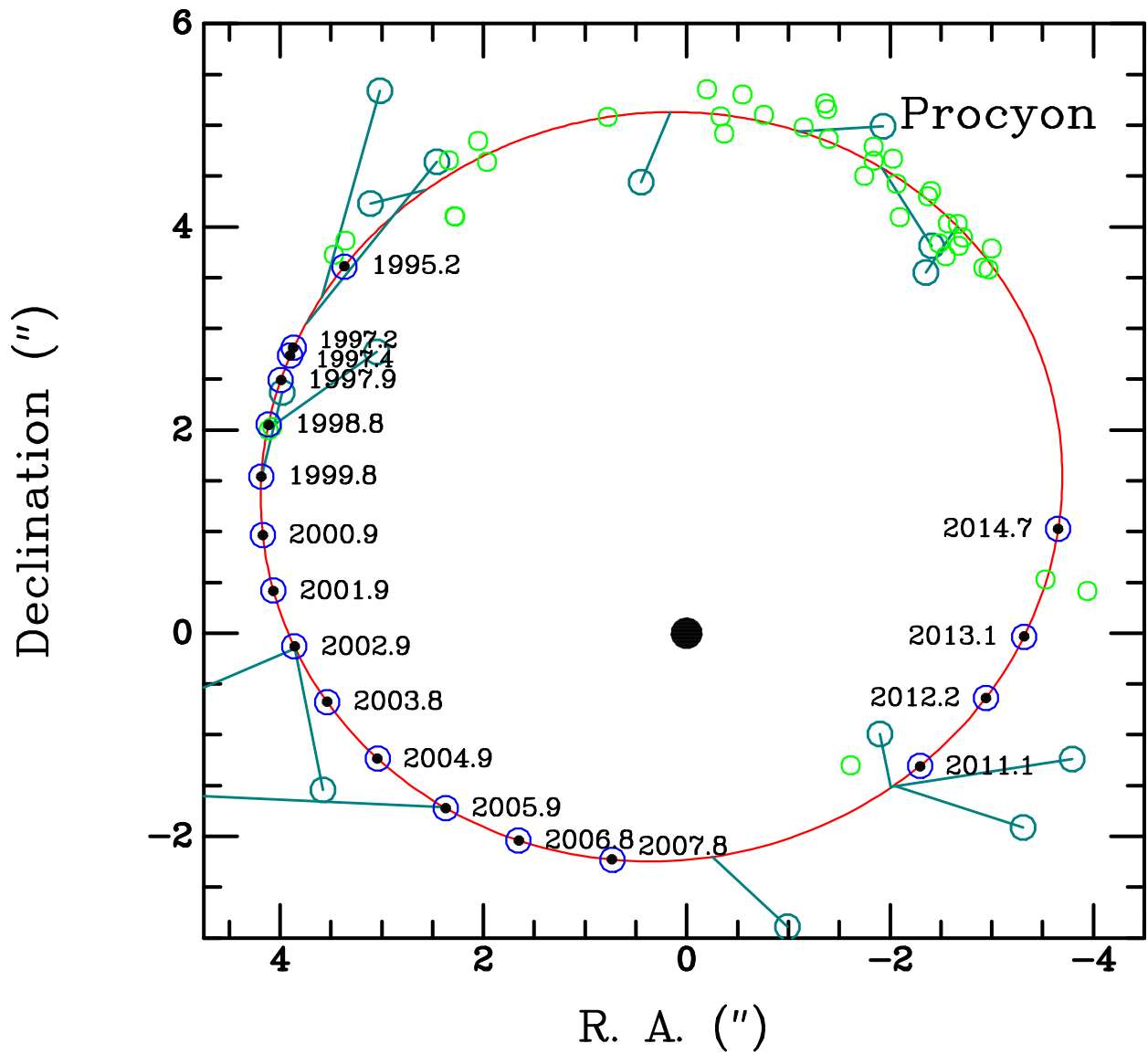}
\end{center}
\vskip-0.4in
\caption{
The relative orbit of Procyon~B\null. \HST\/ observations from
Table~3 are plotted as {\it small filled black circles}. The culled
ground-based observations from Table~7 are shown as {\it open green circles}.
Ground-based measurements that were rejected from our solution are
plotted as {\it open turquoise circles}, connected by straight lines
to their predicted locations.
The {\it solid red curve\/} is our fit to the visual orbit, using the elements
listed in Table~8. {\it Open blue circles\/} mark the positions predicted from
our orbital elements at the dates of the \HST\/ observations, indicated in the
labels.
}
\end{figure}

\begin{figure}
\begin{center}
\includegraphics[width=4.5in]{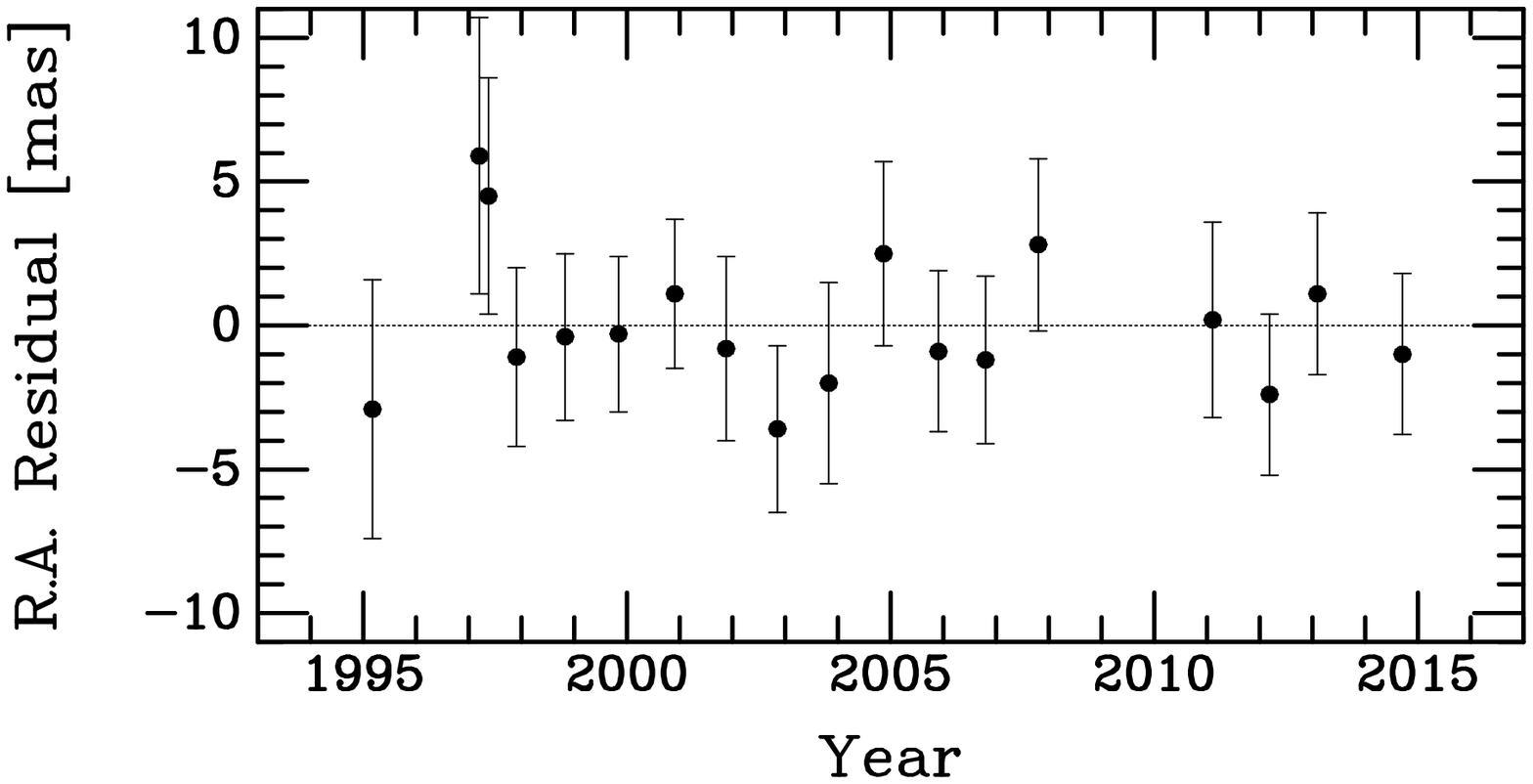}
\vskip 0.2in
\includegraphics[width=4.5in]{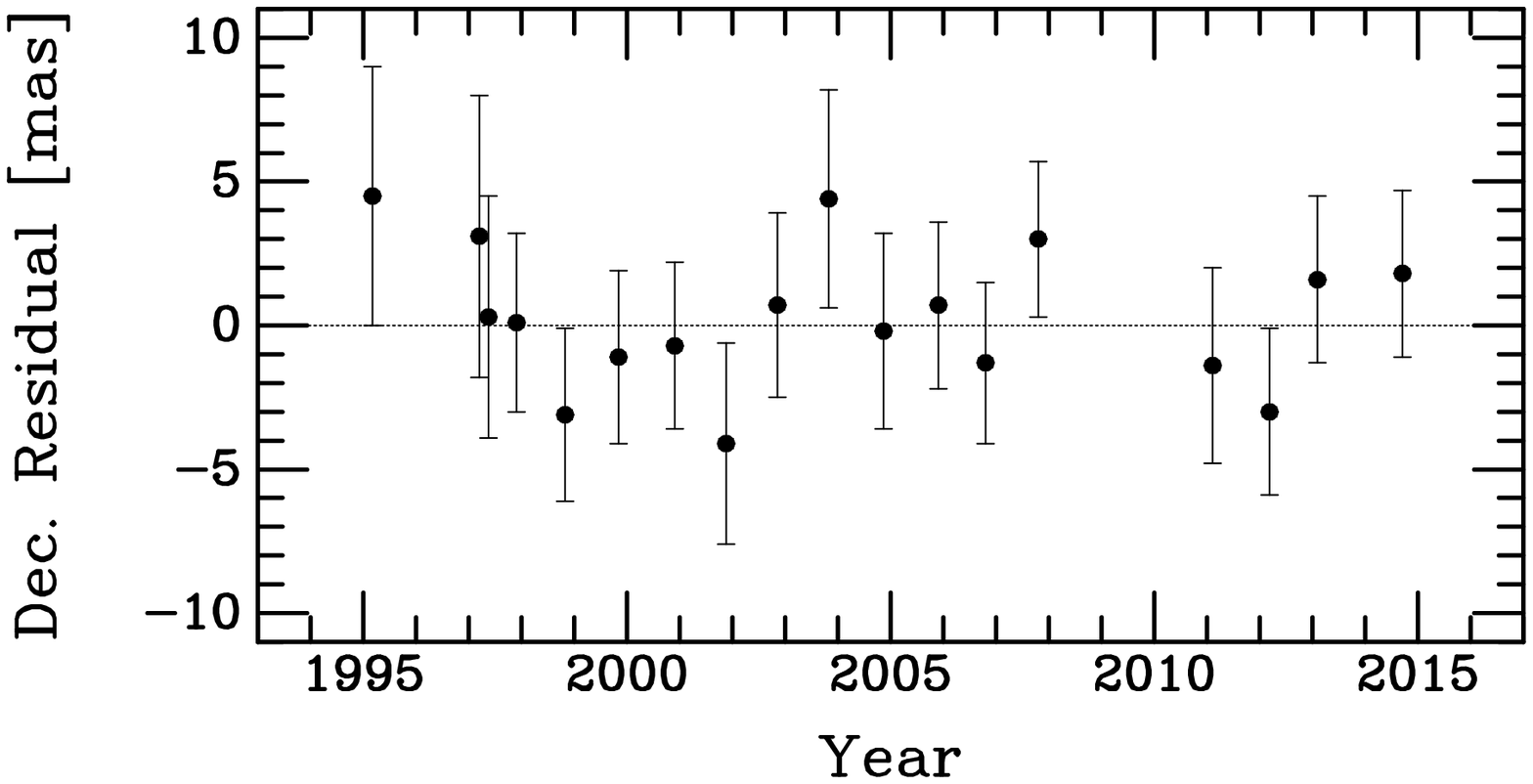}
\end{center}
\vskip-0.2in
\caption{
Residuals (in milliarcseconds) between the right-ascension ({\bf top panel}) and
declination ({\bf bottom panel}) position offsets of Procyon~B from Procyon~A
observed with \HST, and the offsets predicted by our adopted orbital elements.
}
\end{figure}

\begin{figure}
\begin{center}
\includegraphics[width=4.5in]{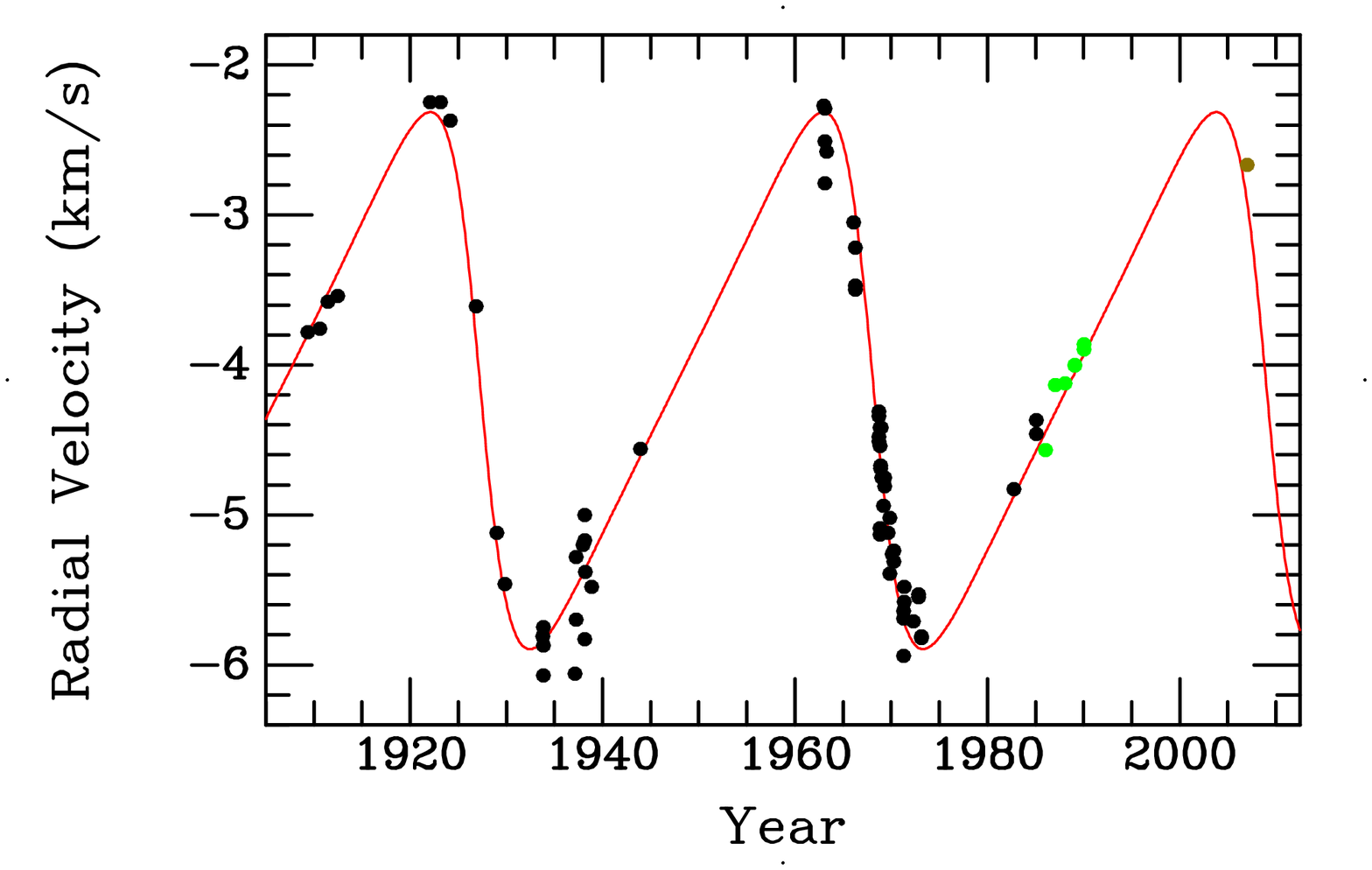}
\includegraphics[width=4.5in]{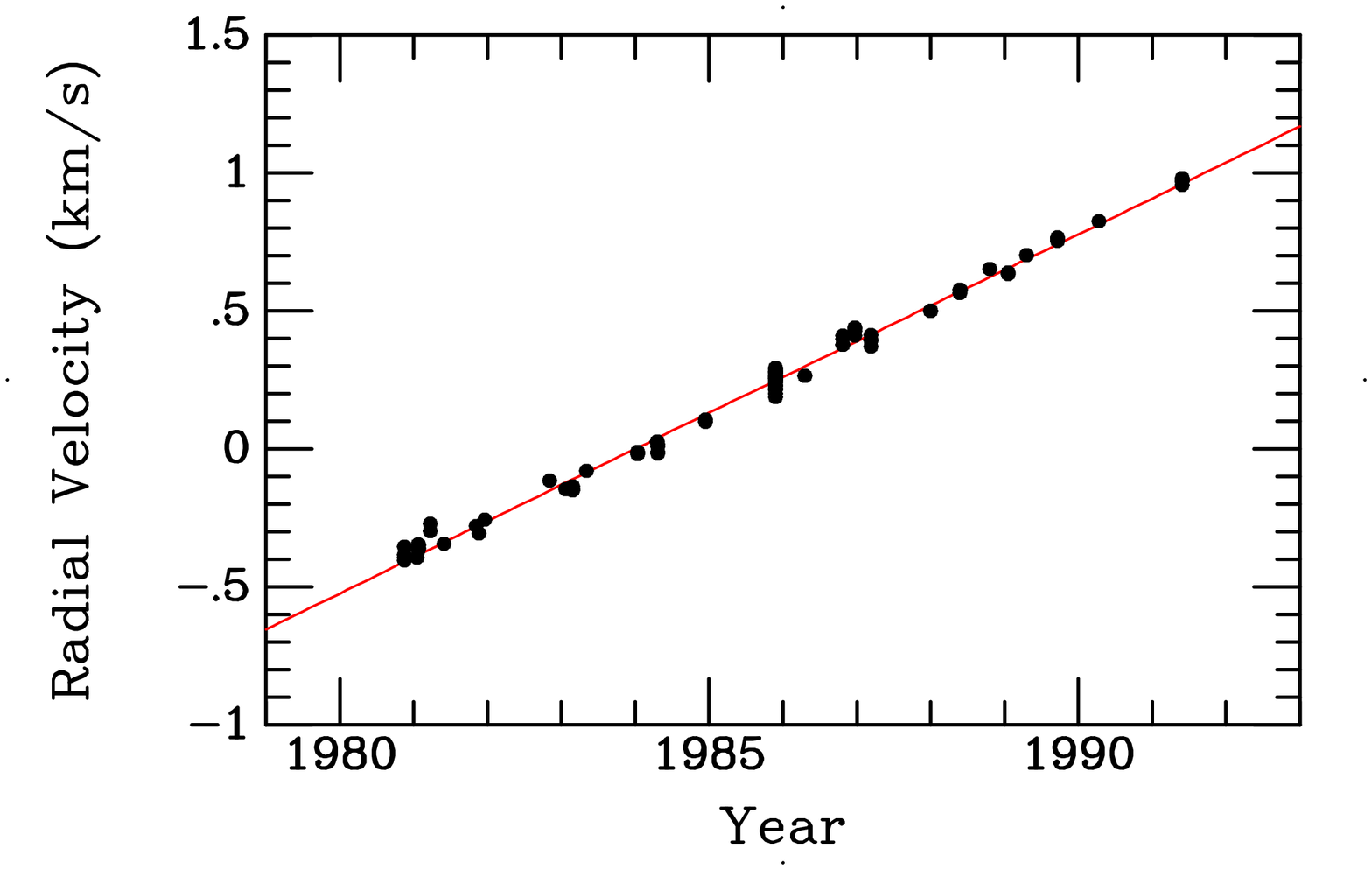}
\end{center}
\vskip-0.3in
\caption{
{\bf Top:} The {\it red line\/} plots the radial-velocity curve of Procyon~A
that is predicted by our orbital elements, the semimajor axis of A's astrometric
motion, and the parallax, with a center-of-mass offset of $-4.115\,\kms$. {\it
Filled black circles\/} are velocity measurements on an absolute scale published
by Irwin et al.\ (1992). {\it Filled green circles\/} are relative velocities
measured by Innis et al.\ (1994), who shifted their zero-point to match that of
Irwin et al. {\it Filled brown circle\/} is an absolute velocity measured by
Mosser et al.\ (2008).
{\bf Bottom:} {\it Filled black circles\/} are precise relative velocity
measurements (Irwin et al.\ 1992). The {\it red line\/} is our predicted
velocity curve, shifted vertically to match the observations in the mean. In
both panels our predictions---based only on astrometry---match the
radial-velocity observations extremely well.
}
\end{figure}

\begin{figure}
\begin{center}
\includegraphics[width=5.5in]{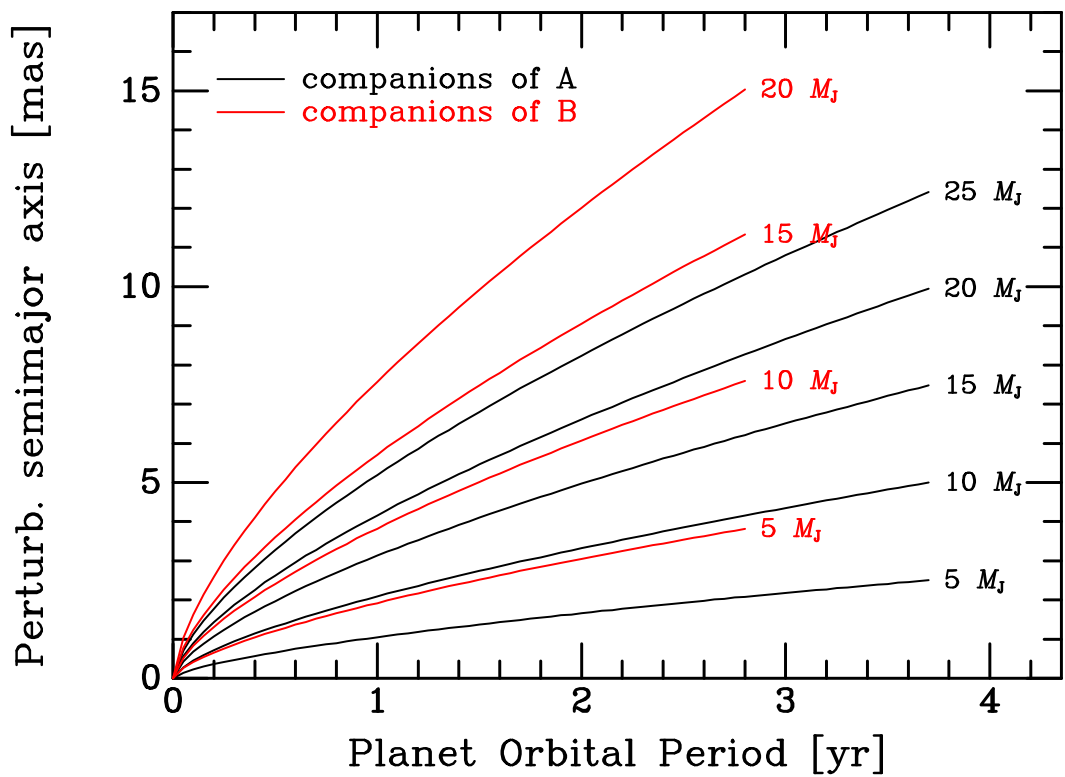}
\end{center}
\vskip-0.2in
\caption{
Astrometric perturbations that would result from planetary companions of
Procyon~A ({\it black curves}) or Procyon~B ({\it red curves}), with the masses
of the perturbers (in units of the mass of Jupiter) indicated in the labels.
Calculations were made for periods up to the orbital-stability limits of planets
with orbital periods of $\sim$3.7~yr (companions of Procyon~A) or $\sim$2.8~yr
(companions of Procyon~B). The $y$-axis is the semimajor axis of the astrometric
perturbation in milliarcseconds.
}
\end{figure}

\begin{figure}
\begin{center}
\includegraphics[width=5.5in]{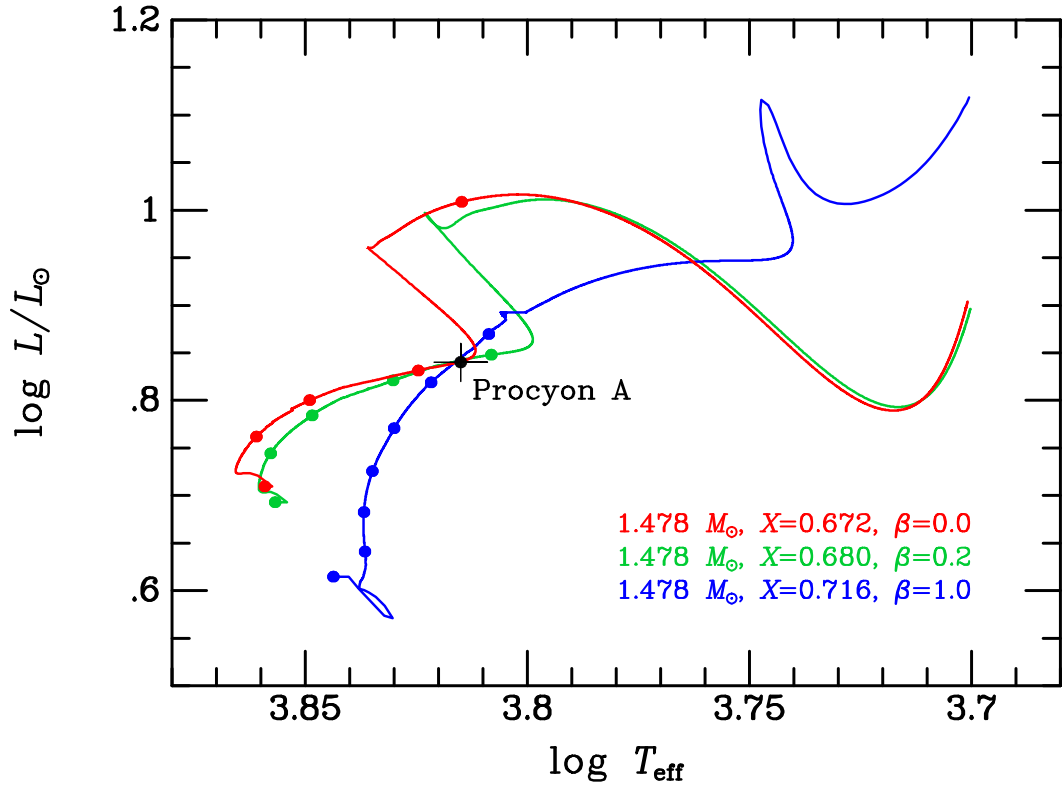}
\end{center}
\vskip-0.2in
\caption{
Theoretical evolutionary tracks in the H-R diagram for stars of $1.478\,M_\odot$
and three different values of the core-overshoot parameter $\beta$: 0.0 (red
curve), 0.2 (green curve), and 1.0 (blue curve).  The hydrogen contents, $X$,
have been adjusted for each track so that it passes through the
location of Procyon~A, marked with a black dot and error bars.  The dots on each
curve are located at ages in steps of 0.5~Gyr, starting at age zero on the ZAMS
at the lower left. For the blue curve, favored by seismic analysis, the age of
Procyon~A is 2.70~Gyr.
}
\end{figure}

\begin{figure}
\begin{center}
\includegraphics[width=4.25in]{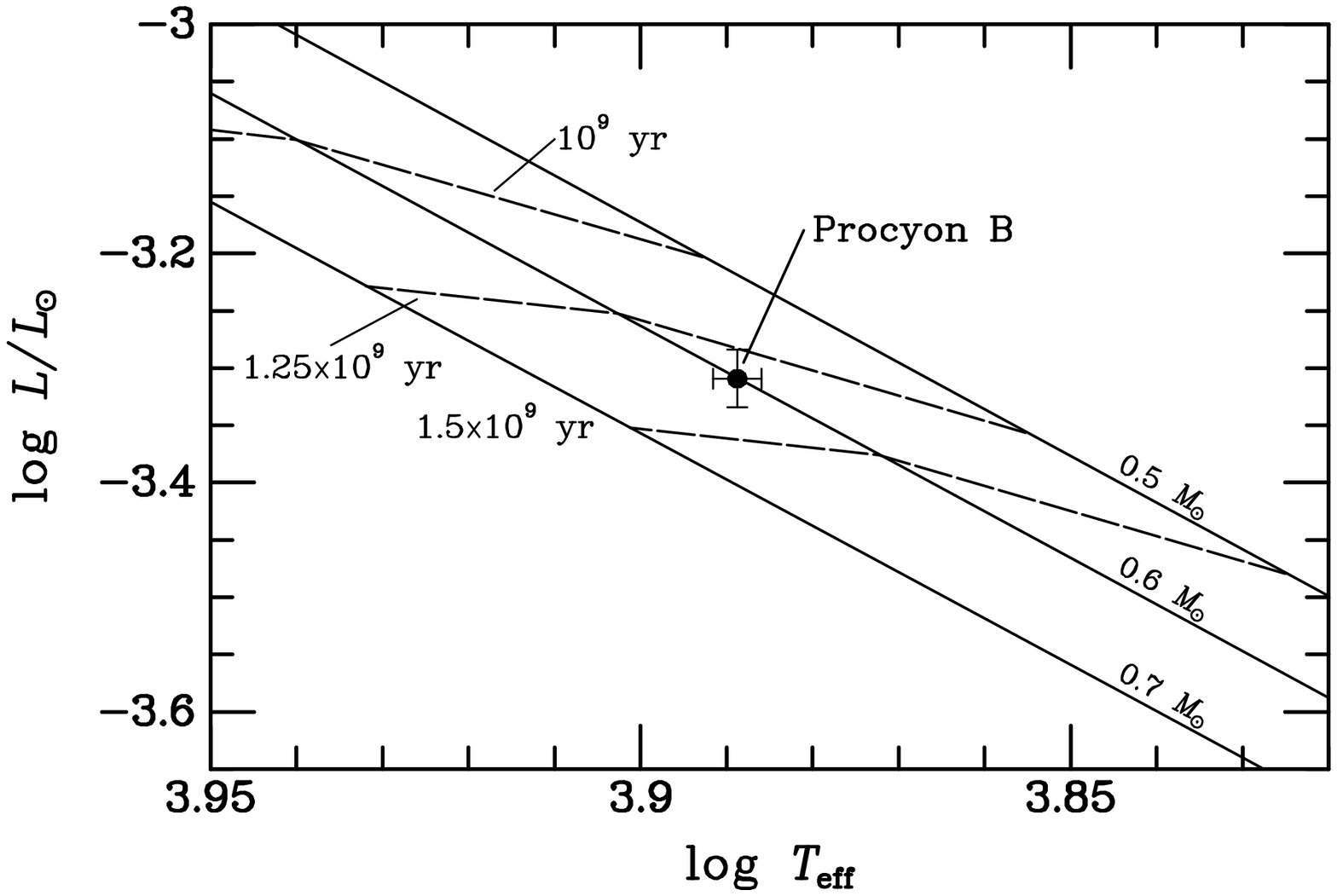}
\vskip 0.2in
\includegraphics[width=4.25in]{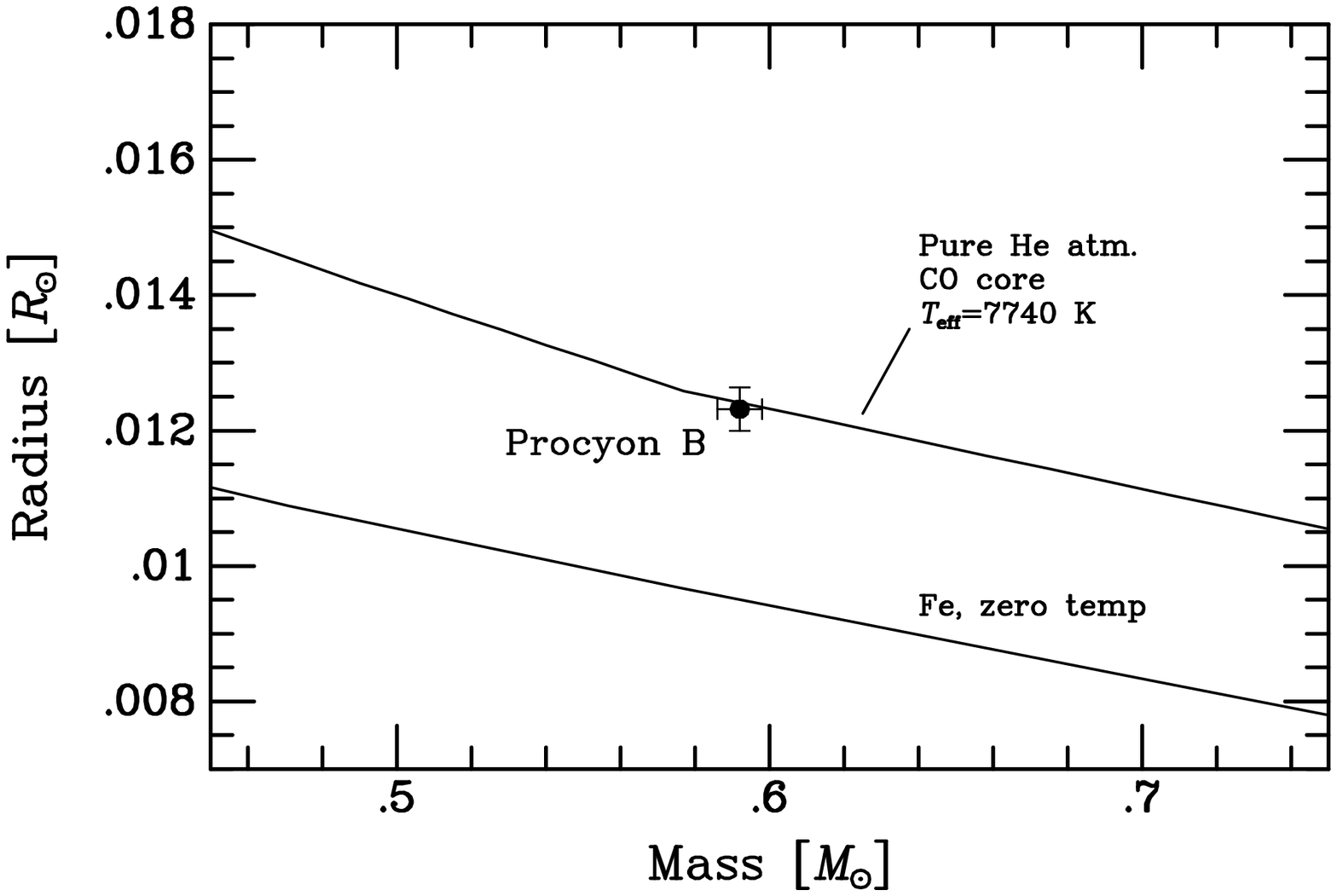}
\end{center}
\vskip-0.2in
\caption{
Comparisons of white-dwarf theory with the observed parameters of
Procyon~B\null. {\bf Top:} Observed position of Procyon~B in the theoretical HR
diagram, compared with Montreal cooling tracks and isochrones for pure
He-atmosphere CO-core white dwarfs of the indicated masses. {\bf Bottom:}
Observed position of Procyon~B in the mass-radius plane, compared with a
theoretical relation for pure He-atmosphere CO white dwarfs of effective
temperature $\Teff=7740$~K, based on the Montreal tracks. Also plotted is the
mass-radius relation for a zero-temperature white dwarf composed of iron (Hamada
\& Salpeter 1961). In both diagrams, the agreement of theory with observations
is excellent, verifying that Procyon~B is a CO-core white dwarf. 
}
\end{figure}


\begin{deluxetable}{lllcccl}
\tablewidth{0 pt}
\tabletypesize{\footnotesize}
\tablecaption{\HST\/ WFPC2 Observing Log for Procyon}
\tablehead{
\colhead{UT Date} &
\colhead{Filter} &
\colhead{Dataset\tablenotemark{a}} &
\colhead{Exposure } &
\colhead{No.\ Pairs} &
\colhead{Proposal} &
\colhead{P. I. } \\
\colhead{} &
\colhead{} &
\colhead{} &
\colhead{Times [s]} &
\colhead{Or Frames\tablenotemark{b}} &
\colhead{ID} &
\colhead{} 
}
\startdata
1995-03-05 & F218W  & u2my0101t & 0.11, 160	 & 1 & 5374  & H. Shipman\tablenotemark{c} \\
1997-03-12 & F1042M & u3mi3702r & 18--200	 & 6 & 6887  & H. Ford   \\
1997-05-17 & F1042M & u3mi3802r & 18--200	 & 5 & 6887  & H. Ford   \\
1997-11-27 & F218W  & u42k0105r & 0.14, 100	 & 2 & 7497  & H.E.B.	 \\
1998-10-29 & F218W  & u42k0801r & 0.12, 100	 & 4 & 7497  & H.E.B.	 \\
1999-11-01 & F218W  & u59h0101r & 0.11, 100--200 & 6 & 8396  & H.E.B.	 \\
2000-11-28 & F218W  & u67h5101r & 0.11, 160--200 & 5 & 8586  & H.E.B.	 \\
2001-11-19 & F218W  & u6iz0101m & 0.11, 160--200 & 5 & 9227  & H.E.B.	 \\
2002-11-08 & F218W  & u8ip0101m & 0.11, 160--200 & 5 & 9332  & H.E.B.	 \\
2003-10-27 & F218W  & u8rm0101m & 0.11, 160--200 & 5 & 9887  & H.E.B.	 \\
2004-11-11 & F218W  & u9290101m & 0.11, 160--200 & 5 & 10112 & H.E.B.	 \\
2005-11-26 & F218W  & u9d30101m & 0.11, 160--200 & 5 & 10481 & H.E.B.	 \\
2006-10-21 & F218W  & u9o50101m & 0.11, 160--200 & 5 & 10914 & H.E.B.	 \\
2007-10-20 & F218W  & ua0p0101m & 0.11, 160--200 & 5 & 11296 & H.E.B.	 \\
\enddata
\tablenotetext{a}{Dataset identifier for the first observation made at each
visit.
}
\tablenotetext{b}{Number of usable F218W short- and long-exposure pairs at same
telescope pointing made during each visit; for the F1042M images, this is the
total number of single frames.
}
\tablenotetext{c}{The short exposure was added to Shipman's program in response
to a Director's Discretionary request submitted by R.L.G.
}
\end{deluxetable}


\begin{deluxetable}{lllccc}
\tablewidth{0 pt}
\tabletypesize{\footnotesize}
\tablecaption{\HST\/ WFC3 Observing Log for Procyon}
\tablehead{
\colhead{UT Date} &
\colhead{Filter} &
\colhead{Dataset\tablenotemark{a}} &
\colhead{Total} &
\colhead{No.} &
\colhead{Proposal} \\
\colhead{} &
\colhead{} &
\colhead{} &
\colhead{Exposure [s]} &
\colhead{Frames\tablenotemark{b}} &
\colhead{ID\tablenotemark{c}} 
}
\startdata
2010-02-24 & F953N & ib7j01010\tablenotemark{d} & 96  & 8  & 11786 \\ 
2010-02-24 & F953N & ib7j01020\tablenotemark{d} & 288 & 8  & 11786 \\ 
2010-02-24 & F953N & ib7j01030\tablenotemark{d} & 96  & 8  & 11786 \\ 
2010-02-24 & F953N & ib7j01040\tablenotemark{d} & 288 & 8  & 11786 \\ 
2011-02-07 & F953N & ibk701010  		& 608 & 16 & 12296 \\ 
2011-02-07 & F953N & ibk701020  		& 608 & 16 & 12296 \\  
2012-03-09 & F953N & ibti01010  		& 608 & 16 & 12673 \\  
2012-03-09 & F953N & ibti01020  		& 608 & 16 & 12673 \\  
2013-02-03 & F953N & ic1k01010  		& 608 & 16 & 13062 \\ 
2013-02-03 & F953N & ic1k01020  		& 608 & 16 & 13062 \\  
2014-09-14 & F953N & ica101010  		& 576 & 16 & 13468 \\  
2014-09-14 & F953N & ica101020  		& 576 & 16 & 13468 \\  
\enddata
\tablenotetext{a}{Dataset identifiers for the drizzle-combined images
obtained during each visit.
}
\tablenotetext{b}{Number of individual dithered sub-exposures contributing to
the drizzle-combined frames.
}
\tablenotetext{c}{H.E.B. was Principal Investigator for all of these
programs.
}
\tablenotetext{d}{The 2010 observations are listed for completeness, but were
not used in our analysis (see text).
}
\end{deluxetable}

\begin{deluxetable}{ccccl}
\tablewidth{0 pt}
\tabletypesize{\footnotesize}
\tablecaption{{\em HST\/} Astrometric Measurements of Procyon B Relative to A}
\tablehead{
\colhead{UT Date} &
\colhead{Besselian} &
\colhead{Separation} &
\colhead{J2000 Position} &
\colhead{Source} \\
\colhead{} &
\colhead{Date} &
\colhead{[arcsec]} &
\colhead{Angle [$^\circ$]} &
\colhead{} 
}
\startdata
1995-03-05& 1995.1745 & $4.9389\pm0.0044$ & $42.977 \pm0.053$ & WFPC2 F218W PSF fit \\ 
1997-03-12& 1997.1958 & $4.7851\pm0.0047$ & $53.997 \pm0.059$ & WFPC2 F1042M spike fit \\
1997-05-17& 1997.3747 & $4.7651\pm0.0040$ & $55.022 \pm0.051$ & WFPC2 F1042M spike fit \\
1997-11-27& 1997.9072 & $4.7058\pm0.0030$ & $58.027 \pm0.039$ & WFPC2 F218W PSF fit \\ 
1998-10-29& 1998.8257 & $4.5973\pm0.0028$ & $63.499 \pm0.038$ & WFPC2 F218W PSF fit \\ 
1999-11-01& 1999.8342 & $4.4583\pm0.0027$ & $69.771 \pm0.039$ & WFPC2 F218W PSF fit \\ 
2000-11-28& 2000.9093 & $4.2809\pm0.0026$ & $76.977 \pm0.039$ & WFPC2 F218W PSF fit \\ 
2001-11-19& 2001.8839 & $4.0859\pm0.0032$ & $84.147 \pm0.049$ & WFPC2 F218W PSF fit \\ 
2002-11-08& 2002.8537 & $3.8584\pm0.0029$ & $91.939 \pm0.047$ & WFPC2 F218W PSF fit \\ 
2003-10-27& 2003.8220 & $3.5988\pm0.0035$ & $100.787\pm0.060$ & WFPC2 F218W PSF fit \\ 
2004-11-11& 2004.8629 & $3.2840\pm0.0032$ & $112.092\pm0.060$ & WFPC2 F218W PSF fit \\ 
2005-11-26& 2005.9040 & $2.9293\pm0.0027$ & $125.956\pm0.058$ & WFPC2 F218W PSF fit \\ 
2006-10-21& 2006.8046 & $2.6266\pm0.0027$ & $140.997\pm0.065$ & WFPC2 F218W PSF fit \\ 
2007-10-20& 2007.8011 & $2.3452\pm0.0027$ & $161.715\pm0.074$ & WFPC2 F218W PSF fit \\ 
\noalign{\smallskip}
2011-02-07& 2011.1040 & $2.6431\pm0.0047$ & $240.339\pm0.105$ & WFC3 F953N PSF fit \\ 
2012-03-09& 2012.1877 & $3.0130\pm0.0040$ & $257.721\pm0.078$ & WFC3 F953N PSF fit \\ 
2013-02-03& 2013.0947 & $3.3154\pm0.0040$ & $269.417\pm0.071$ & WFC3 F953N PSF fit \\ 
2014-09-14& 2014.7038 & $3.7986\pm0.0040$ & $285.719\pm0.062$ & WFC3 F953N PSF fit \\ 
\noalign{\smallskip}
2011-02-07& 2011.1040 & $2.6381\pm0.0048$ & $240.205\pm0.106$ & WFC3 F953N spike fit \\
2012-03-09& 2012.1877 & $3.0144\pm0.0040$ & $257.752\pm0.078$ & WFC3 F953N spike fit \\
2013-02-03& 2013.0947 & $3.3166\pm0.0040$ & $269.447\pm0.071$ & WFC3 F953N spike fit \\
2014-09-14& 2014.7038 & $3.7946\pm0.0040$ & $285.723\pm0.063$ & WFC3 F953N spike fit \\
\noalign{\smallskip}
2011-02-07& 2011.1040 & $2.6406\pm0.0034$ & $240.272\pm0.074$ & WFC3 F953N average \\
2012-03-09& 2012.1877 & $3.0137\pm0.0028$ & $257.737\pm0.055$ & WFC3 F953N average \\
2013-02-03& 2013.0947 & $3.3160\pm0.0028$ & $269.432\pm0.050$ & WFC3 F953N average \\
2014-09-14& 2014.7038 & $3.7966\pm0.0028$ & $285.721\pm0.044$ & WFC3 F953N average \\
\enddata
\end{deluxetable}


\begin{deluxetable}{lllcccl}
\tablewidth{0 pt}
\tabletypesize{\footnotesize}
\tablecaption{\HST\/ WFPC2 F1042M Frames used for PSF Studies\tablenotemark{a}}
\tablehead{
\colhead{Target} &
\colhead{UT Date} &
\colhead{Dataset\tablenotemark{b}} &
\colhead{Exposure} &
\colhead{No.} &
\colhead{Proposal} &
\colhead{P. I.}\\
\colhead{} &
\colhead{} &
\colhead{} &
\colhead{Times [s]} &
\colhead{Frames} &
\colhead{ID} &
\colhead{}  
}
\startdata
Sirius  & 1997-03-19 & u3mi1503r & 12--100    & 4  & 6887  & H. Ford \\
Sirius  & 1997-05-18 & u3mi1603m & 12--100    & 4  & 6887  & H. Ford \\
Sirius  & 2001-10-27 & u6gb0202m & 4--35      & 10 & 9072  & H.E.B. \\ 
Sirius  & 2002-05-10 & u6gb0306m & 4--60      & 10 & 9072  & H.E.B. \\ 
Sirius  & 2002-10-20 & u8if0206m & 8--60      & 10 & 9334  & H.E.B. \\ 
Sirius  & 2003-04-18 & u8if0306m & 8--60      & 10 & 9334  & H.E.B. \\ 
Sirius  & 2003-10-15 & u8tp0206m & 8--60      & 10 & 9964  & H.E.B. \\ 
Sirius  & 2004-08-15 & u8tp0301m & 8--60      & 12 & 9964  & H.E.B. \\ 
Sirius  & 2005-04-20 & u8tp0601m & 8--60      & 12 & 9964  & H.E.B. \\ 
Sirius  & 2006-01-15 & u9bv0101m & 8--60      & 12 & 10619 & H.E.B. \\ 
Sirius  & 2006-12-27 & u9o60101m & 8--60      & 13 & 10990 & H.E.B. \\ 
Sirius  & 2008-01-03 & u9z80101m & 8--60      & 12 & 11290 & H.E.B. \\ 
109 Vir & 2008-04-06 & ub080101m & 0.23--600  & 3  & 11509 & R.L.G. \\ 
\enddata
\tablenotetext{a}{For PSF definition to be used in centroiding B, only the
Sirius exposures longer than 30~s were used, along with the 0.23-s exposures of
109~Vir.
}
\tablenotetext{b}{Dataset identifier for the first observation made at each
visit.
}
\end{deluxetable}

\begin{deluxetable}{lllccc}
\tablewidth{0 pt}
\tabletypesize{\footnotesize}
\tablecaption{\HST\/ WFC3 F953N Frames used for PSF Studies}
\tablehead{
\colhead{Target} &
\colhead{UT Date} &
\colhead{Dataset} &
\colhead{Total} &
\colhead{No.} &
\colhead{Proposal} \\
\colhead{} &
\colhead{} &
\colhead{} &
\colhead{Exposure [s]} &
\colhead{Exposures\tablenotemark{a}} &
\colhead{ID\tablenotemark{b}} 
}
\startdata
Sirius   & 2010-09-02 & ibk703010 & 48   & 8 & 12296  \\  
Sirius   & 2010-09-02 & ibk703020 & 96   & 8 & 12296  \\  
Sirius   & 2010-09-02 & ibk703030 & 24   & 4 & 12296  \\  
Sirius   & 2010-09-02 & ibk703040 & 96   & 8 & 12296  \\  
Sirius   & 2011-10-01 & ibti03010 & 48   & 8 & 12673  \\  
Sirius   & 2011-10-01 & ibti03020 & 96   & 8 & 12673  \\  
Sirius   & 2011-10-01 & ibti03030 & 24   & 4 & 12673  \\  
Sirius   & 2011-10-01 & ibti03040 & 96   & 8 & 12673  \\  
HD 23886 & 2012-02-17 & ibs001010 & 20   & 4 & 12598  \\  
HD 23886 & 2012-02-17 & ibs001020 & 2524 & 4 & 12598  \\ 
Sirius   & 2012-09-26 & ic1k03010 & 48   & 8 & 13062  \\  
Sirius   & 2012-09-26 & ic1k03020 & 96   & 8 & 13062  \\  
Sirius   & 2012-09-26 & ic1k03030 & 24   & 4 & 13062  \\  
Sirius   & 2012-09-26 & ic1k03040 & 96   & 8 & 13062  \\  
Sirius   & 2014-03-31 & ica103010 & 48   & 8 & 13468  \\  
Sirius   & 2014-03-31 & ica103020 & 96   & 8 & 13468  \\  
Sirius   & 2014-03-31 & ica103030 & 24   & 4 & 13468  \\  
Sirius   & 2014-03-31 & ica103040 & 96   & 8 & 13468  \\  
\enddata
\tablenotetext{a}{Number of individual frames used to create the listed
drizzle-combined images.
}
\tablenotetext{b}{H.E.B. was Principal Investigator for all of these
programs.
}
\end{deluxetable}


\begin{deluxetable}{lllllrllll}
\tabletypesize{\scriptsize}
\tablecaption{Adjustments to Measures Compiled by Strand and WDS}
\tablewidth{0pt}
\tablehead{
\colhead{Date} & \colhead{Sep.} & \colhead{PA\tablenotemark{a}} & \colhead{Observer/} &
 \colhead{Catalog} & \colhead{Action} & \colhead{Date} & \colhead{Sep.} & 
  \colhead{PA\tablenotemark{a}} & \colhead{Ref.\ Code\tablenotemark{b}} \\
\colhead{} & \colhead{[$''$]} & \colhead{[$^\circ$]} & \colhead{Ref.\ Code\tablenotemark{b}} & 
 \colhead{} & \colhead{} & \colhead{} & \colhead{[$''$]} & \colhead{[$^\circ$]} & \colhead{} 
} 
\startdata
1897.83 & 4.82 & 329.1 & Boothroyd   & Strand/WDS & Replaced by & 1898.129 & 4.78 & 327.52 & Boo1898    \\
1897.83 & 4.80 & 327.9 & See1898c    & Strand/WDS & Replaced by & 1898.189 & 4.57 & 327.12 & See1898e   \\
1898.21 & 4.82 & 325.3 & A\_\_1899b  & Strand/WDS & Replaced by & 1898.050 & 4.75 & 325.12 & A\_\_1914d \\
1898.76 & 4.97 & 330.9 & A\_\_1899b  &        WDS & Replaced by & 1898.880 & 4.97 & 331.11 & A\_\_1914d \\
1899.25 & 4.99 & 329.6 & A\_\_1900d  &        WDS & Replaced by & 1898.880 & 4.97 & 331.11 & A\_\_1914d \\
1902.72 & 5.33 & 351.1 & Aitken      & Strand	  & Replaced by & 1902.241 & 5.34 & 345.00 & A\_\_1914d \\
        &      &       &	     &  	  &		& 1902.960 & 5.33 & 354.09 & A\_\_1914d \\
1905.14 & 5.14 &   6.7 & Aitken      & Strand	  & Replaced by & 1905.570 & 5.14 &   8.68 & A\_\_1914d \\
1910.10 & 5.21 &  24.8 & Barnard     & Strand	  & Replaced by & 1910.025 & 5.21 &  26.71 & Bar1912    \\
1928.98 & 2.68 & 237.9 & van den Bos & Strand	  & Replaced by & 1928.824 & 2.07 & 231.06 & B\_\_1929a \\
        &      &       &	     &  	  &		& 1929.041 & 2.14 & 242.56 & B\_\_1929a \\
        &      &       &	     &  	  &		& 1929.079 & 3.82 & 240.06 & B\_\_1929a \\
\enddata 
\tablenotetext{a}{Position angles are given for J2000 equinox.}
\tablenotetext{b}{The reference code as defined in the Washington Double Star
Catalog, http://ad.usno.navy.mil\slash wds\slash Webtextfiles/wdsnewref.txt. All
of the ``replaced by'' values are taken from the WDS catalog.}
\label{tab.remove}
\end{deluxetable}

\begin{deluxetable}{lcclc}
\tabletypesize{\footnotesize}
\tablecaption{Ground-Based Measurements of Procyon Used in Orbit Fit}
\tablewidth{0pt}
\tablehead{
\colhead{Besselian}    & \colhead{Sep.} & \colhead{J2000 Position} &
\colhead{Ref.\ Code\tablenotemark{a}} & \colhead{Method\tablenotemark{b}} \\ 
\colhead{Date} & \colhead{[$''$]} & \colhead{Angle [$^\circ$]} & \colhead{}         &
\colhead{}}
\startdata
1896.930  &  4.63  &  320.92  &  Shb1897a   & Ma \\
1897.000  &  4.83  &  321.62  &  A\_\_1914d & Ma \\
1897.160  &  4.65  &  320.32  &  Hu\_1898   & Ma \\
1897.821  &  4.66  &  324.92  &  Shb1897b   & Ma \\
1898.050  &  4.75  &  325.12  &  A\_\_1914d & Ma \\
1898.129  &  4.78  &  327.52  &  Boo1898    & Ma \\
1898.189  &  4.57  &  327.12  &  See1898e   & Ma \\
1898.213  &  4.83  &  326.52  &  Bar1898a   & Ma \\
1898.240\tablenotemark{c}  &    4.26  &  326.50  &  Lewis       &  Ma  \\
1898.282  &  4.50  &  325.52  &  Hu\_1903b  & Ma \\
1898.880  &  4.97  &  331.11  &  A\_\_1914d & Ma \\
1899.073  &  4.91  &  331.11  &  Bar1899    & Ma \\
1899.960  &  4.88  &  335.01  &  A\_\_1914d & Ma \\
1900.055  &  5.09  &  336.54  &  Bar1900c   & Ma \\
1900.236  &  4.83  &  338.81  &  L\_\_1900  & Ma \\
1900.252\tablenotemark{c}  &	4.51  &  327.71  &  See1911	&  Ma  \\
1900.295  &  4.60  &  332.91  &  See1900d   & Ma \\
1901.200  &  5.13  &  339.00  &  A\_\_1901b & Ma \\
1901.300  &  5.00  &  338.40  &  See1911    & Ma \\
1901.883  &  5.06  &  343.99  &  Bar1903a   & Ma \\
1902.214  &  5.39  &  345.40  &  L\_\_1902a & Ma \\
1902.241  &  5.34  &  345.00  &  A\_\_1914d & Ma \\
1902.241  &  5.11  &  347.00  &  Hu\_1903b  & Ma \\
1902.253\tablenotemark{c}  &	5.35  &  338.90  &  See1911	&  Ma  \\
1902.960  &  5.33  &  354.09  &  A\_\_1914d & Ma \\
1903.154  &  5.16  &  351.52  &  Bar1903a   & Ma \\
1904.294  &  4.93  &  355.69  &  Bow1904a   & Ma \\
1904.795  &  5.36  &  357.87  &  Bar1909b   & Ma \\
1905.170\tablenotemark{c}  &	4.46  &    5.78  &  L\_\_1905	&  Ma  \\
1905.570  &  5.14  &	8.68  &  A\_\_1914d & Ma \\
1909.162  &  5.26  &   22.97  &  Bar1909b   & Ma \\
1909.298  &  5.04  &   22.96  &  Bow1909    & Ma \\
1910.025  &  5.21  &   26.71  &  Bar1912    & Ma \\
1911.060  &  4.70  &   29.10  &  J\_\_1917c & Ma \\
1911.069  &  4.69  &   29.05  &  J\_\_1911e & Ma \\
1913.162  &  5.09  &   43.00  &  Bar1913    & Ma \\
1914.300\tablenotemark{c}  &	6.14  &   29.50  &  Bowyer	&  Ma  \\
1914.939\tablenotemark{c}  &	5.25  &   27.93  &  J\_\_1917c  &  Ma  \\ 
1917.241\tablenotemark{c}  &	4.12  &   47.82  &  J\_\_1917c  &  Ma  \\	 
1918.220\tablenotemark{c}  &	4.63  &   59.22  &  J\_\_1918b  &  Ma  \\	 
1921.214\tablenotemark{c}  &	5.61  &   98.90  &  StG1962a	&  Ma  \\ 
1924.190\tablenotemark{c}  &	5.45  &  106.88  &  Dic1962	&  Ma  \\ 
1927.106\tablenotemark{c}  &	3.06  &  198.97  &  B\_\_1929a  &  Ma  \\ 
1928.824  &  2.07  &  231.06  &  B\_\_1929a & Ma \\
1929.041\tablenotemark{c}  &	2.14  &  242.56  &  B\_\_1929a  &  Ma  \\ 
1929.060\tablenotemark{c}  &	3.99  &  251.96  &  Fin1934b	&  Ma  \\	 
1929.079\tablenotemark{c}  &	3.82  &  240.06  &  B\_\_1929a  &  Ma  \\ 
1932.272  &  3.57  &  278.54  &  B\_\_1932b & Ma \\
1932.277  &  3.96  &  276.04  &  Fin1934b   & Ma \\
1957.840  &  4.554 &   63.51  &  vAb1958    & Po \\
1957.853  &  4.573 &   64.06  &  The1975    & Po \\
1962.000\tablenotemark{c}  &	3.90  &  113.19  &  B\_\_1962d  &  Ma  \\ 
1986.254  &  5.10  &  356.27  &  Wor1989    & Ma \\
1992.720\tablenotemark{c}  &	5.25  &   36.30  &  WGA1994	&  AO  \\
1995.090  &  5.12  &   41.00  &  Grr2000    & CCD \\
\enddata 
\tablenotetext{a}{The reference code as defined in the Washington Double Star
Catalog, http://ad.usno.navy.mil\slash wds\slash Webtextfiles/wdsnewref.txt}
\tablenotetext{b}{WDS method codes are: Ma (micrometer), Po (photography), CCD
(CCD imaging), AO (adaptive optics).}
\tablenotetext{c}{Observation rejected from our orbital solution.}
\label{tab.orbmeas}
\end{deluxetable} 

\begin{deluxetable}{ll}
\tablewidth{0 pt}
\tablecaption{Elements of Relative Visual Orbit of Procyon}
\tablehead{
\colhead{Element} &
\colhead{Value} 
}
\startdata
Orbital period, $P$ [yr]                &   40.840   $\pm$ 0.022   \\
Semimajor axis, $a$ [arcsec]            &    4.3075  $\pm$ 0.0016  \\
Inclination, $i$ [deg]                  &   31.408   $\pm$ 0.050   \\
Position angle of node, $\Omega$ [deg]  &  100.683   $\pm$ 0.095   \\
Date of periastron passage, $T_0$ [yr]  & 1968.076   $\pm$ 0.023   \\
Eccentricity, $e$                       &    0.39785 $\pm$ 0.00025 \\
Longitude of periastron, $\omega$ [deg] &   89.23    $\pm$ 0.11    \\
\enddata
\end{deluxetable}

\begin{deluxetable}{lll}
\tablewidth{0 pt}
\tablecaption{Parallax and Semimajor Axis for Procyon A}
\tablehead{
\colhead{Source} & 
\colhead{Value}  &
\colhead{Reference} 
}
\startdata
\multispan3{\hfil Absolute Parallax, $\pi$ [arcsec] \hfil} \\
\noalign{\vskip-0.25in}\\
USNO plates & $ 0.2832 \pm 0.0015 $ & Girard et al.\ (2000)  \\
\Hipp\/     & $ 0.2846 \pm 0.0013 $ & van Leeuwen (2007)     \\
MAP	    & \underline{$ 0.2860 \pm 0.0010 $} & Gatewood \& Han (2006) \\
\noalign{\vskip-0.275in} \\
Weighted mean & $ 0.2850 \pm 0.0007 $ &        \\
\noalign{\vskip-0.2in} \\
\multispan3{\hfil Semimajor Axis, $a_A$ [arcsec] \hfil} \\
$\sim$600 exposures & $ 1.232  \pm 0.008  $ & Girard et al.\ (2000) \\
\enddata
\end{deluxetable}

\begin{deluxetable}{ll}
\tablewidth{0 pt}
\tablecaption{Dynamical Masses for Procyon System}
\tablehead{
\colhead{Mass} & 
\colhead{Value}  
}
\startdata
Total mass, $M_A+M_B$ & $2.070 \pm 0.016 \, M_\odot$ \\
$M_A$                 & $1.478 \pm 0.012 \, M_\odot$ \\
$M_B$                 & $0.592 \pm 0.006 \, M_\odot$ \\
\enddata
\end{deluxetable}

\begin{deluxetable}{lllcc}
\tablewidth{0 pt}
\tablecaption{Error Budgets for Procyon System Dynamical Masses}
\tablehead{
\colhead{Quantity} &
\colhead{Value} &
\colhead{Uncertainty} &
\colhead{$\sigma(M_A)$ [$M_\odot$] } &
\colhead{$\sigma(M_B)$ [$M_\odot$] }
}
\startdata
Absolute Parallax, $\pi$    & 0.2850  & $\pm$0.0007 arcsec & 0.0109 & 0.0044 \\
Semimajor axis, $a$         & 4.3075  & $\pm$0.0016 arcsec & 0.0019 & 0.0004 \\
Semimajor axis for A, $a_A$ & 1.232   & $\pm$0.008  arcsec & 0.0038 & 0.0038 \\
Period, $P$                 & 40.840  & $\pm$0.022  yr     & 0.0016 & 0.0006 \\
\noalign{\vskip0.1in}
Combined mass uncertainty   &         &                    & 0.012  & 0.006 \\
\enddata
\end{deluxetable}

\end{document}